\documentclass[12pt,preprint]{aastex}
\usepackage{epsfig}
\usepackage{amsmath}

\def\hb{H$\beta$}
\def\mgii{Mg {\sc ii}~}
\def\civ{C {\sc iv}~}

\usepackage{epsfig}
\usepackage{graphicx}
\usepackage{ifthen}
\usepackage{amssymb}
\usepackage{subfigure}

\begin{document}
\title{The jet power, radio loudness and black hole mass in radio loud AGNs }

\author{Yi Liu \altaffilmark{1,2,3}, Dong Rong Jiang\altaffilmark{1,2}, Min Feng Gu\altaffilmark{1,2}}
\altaffiltext{1}{Shanghai Astronomical Observatory, Chinese
Academy
of Sciences, Shanghai 200030, China\\
E-mail: yliu@center.shao.ac.cn }

\altaffiltext{2}{National Astronomical Observatories, Chinese
Academy of Sciences, Beijing 100012, China }
\altaffiltext{3}{Graduate School of the Chinese Academy of
Sciences, BeiJing 100039, China }

\begin{abstract}
The jet formation is thought to be closely connected with the mass
of central supermassive black hole in Active Galactic Nuclei. The
radio luminosity commonly used in investigating this issue is
merely an indirect measure of the energy transported through the
jets from the central engine, and severely Doppler boosted in
core-dominated radio quasars. In this work, we investigate the
relationship between the jet power and black hole mass, by
estimating the jet power using extrapolated extended 151 MHz flux
density from the VLA 5 GHz extended radio emission, for a sample
of 146 radio loud quasars complied from literature. After removing
the effect of relativistic beaming in the radio and optical
emission, we find a significant intrinsic correlation between the
jet power and black hole mass. It strongly implies that the jet
power, so as jet formation, is closely connected with the black
hole mass.
To eliminate the beaming effect in the conventional radio
loudness, we define a new radio loudness as the ratio of the radio
extended luminosity to the optical luminosity estimated from the
broad line luminosity.
In a tentatively combined sample of radio quiet with our radio
loud quasars, the apparent gap around the conventional radio
loudness R=10 is not prominent for the new-defined radio loudness.
In this combined sample, we find a significant correlation between
the black hole mass and new-defined radio loudness.

\end{abstract}


\section{INTRODUCTION}
Relativistic jets observed in radio loud AGNs as an extreme
phenomenon, congregated most attention in the past decades. The
properties of relativistic radio jets are thought to be closely
connected with the properties of both accretion disk and black
hole in active galactic nuclei. However, the origin and formation
of relativistic jets are still the unsolved mystery lying in
astrophysics. The currently most favored models of the formation
of the jet are Blandford-Znajek and Blandford-Payne mechanisms
(Blandford \& Znajek 1977; Blandford \& Payne 1982). In both
mechanisms, the accretion generated the power and extracted from
the disk or black hole rotational energy and converted into the
kinetic power of the jet. Moreover, the accretion process upon the
central supermassive black hole is believed to be responsible for
the activity of AGNs. It is thus conceivable that the radio
activity is connected with the central black hole.


Since the correlation of black hole mass with radio luminosity was
first suggested for a handful of galaxies (Franceschini,
Vercellone \& Fabian 1998), the issue of a possible dependence of
radio activity on black hole mass has recently been the subject of
many scientific debates (McLure et al. 1999; Lacy et al. 2001;
Jarvis \& McLure 2002; McLure \& Jarvis 2004). Several studies
claimed that the radio luminosity is tightly connected with the
black hole mass (e.g. Laor 2000; Lacy et al. 2001; McLure et al.
2004) in the different samples.
Recently, using a sample of about 6000 quasars from the SDSS,
McLure \& Jarvis (2004) proposed that both the radio luminosity
and radio loudness are strongly correlated with the black hole
mass, although the range in radio luminosity at a given black hole
mass is several orders of magnitude. However, several authors have
claimed that no strong link exists between the black hole mass and
radio luminosity in AGNs (e.g. Oshlack, Webster \& Whiting 2002;
Woo \& Urry 2002a; Ho 2002; Urry 2003; Snellen et al. 2003).
Woo \& Urry (2002a) showed that the black hole mass ranges are not
different between radio loud and radio quiet samples with 377
AGNs, and the strong correlation between black hole mass and radio
luminosity is not found. Moreover, Oshlack et al. (2002) found no
indication of correlation between the radio luminosity and black
hole mass for a sample of radio-selected, flat-spectrum quasars
(see Jarvis \& McLure 2002 for a different interpretation).

In spite of these widely contrasting results, it is worth noting
that the radio emission is commonly used in the above mentioned
works, however, this can be problematic for radio loud AGNs. Due
to the significant Doppler enhancement in the relativistic jets,
the total radio luminosity is only a poor indicator of intrinsic
jet power.
Specifically, the majority of core-dominated blazar-like quasars
have incredibly strong 5 GHz flux density from emission on the
subkiloparsec scale, yet they have weak or moderate radio lobe
emission (Punsly 1995). Moreover, the radio emission dissipated in
the jet only represents the part of the jet power, and most of the
energy in the jets is not radiated away but is transported to the
lobes. Thus, the radio emission is an indirect measurement of the
energy transported through the jets from the central engine.
Physically, a far better way is to investigate the relationship
between the jet power and black hole mass. Usually, the jet power
as a fundamental radio parameter to indicate the energy
transported through the radio jet from the central engine, can not
be readily obtained. Nevertheless, it can be estimated from
studying the isotropic properties of the material ejected from the
ends of the jets in the radio lobes. Rawlings \& Saunders (1991)
estimated $\rm Q_{jet}$, the bulk kinetic power as a token of
total starting jet kinetic power, under the main assumption that
the electrons and the magnetic field make an equal contribution to
the total energy density. Then from a total lobe energy E, an
efficiency $\rm \eta$ that allows for work done on the external
medium, and a lobe age T, the $\rm Q_{jet}$ can be derived by
using $\rm Q=E/T\eta$. Recently, Willott et al. (1999) presented a
sophisticated calculation of the jet power utilizing the optically
thin flux density from the lobes measured at 151 MHz, which
incorporates the deviations from the minimum-energy estimates in a
multiplicative factor $f$ that represents the small departures
from minimum energy, geometric effects, filling factors, protonic
contributions, and the low-frequency cutoff.

More recently, motivated by recent X-ray observations, Punsly
(2005) presented a theoretical derivation of an estimate for a
radio source jet power, by assuming that most of the energy in the
lobes is in plasma thermal energy with a negligible contribution
from magnetic energy (not equipartition), and computing the
elapsed time from spectral ageing. The basic idea is that lobe
expansion is dictated by the internal dynamics of the lobes and
the physical state of the enveloping extragalactic gas. The
expression yields jet powers that are quantitatively similar (to
within a factor of 2) to the more sophisticated empirical relation
of Willott et al. (1999) and Blundell \& Rawlings (2000), in spite
of the fact that U, the energy stored in the lobes, and T, the
elapsed time, are determined from completely different methods and
assumptions. The formula allows one to estimate the jet power
using the measurement of the optically thin radio lobe emission in
quasars and radio galaxies. The close agreement of the two
independent expressions makes them robust estimators of jet power.

Usually, the radio loudness can be used to indicate the radio
properties, and to distinguish radio quiet and radio loud
populations. It is conventionally defined as $\rm R\equiv
L_{\nu5GHz}/L_{\nu4400}$, and the boundary between the two
populations was set at $\rm R=10$ (Visnovsky et al. 1992; Stocke
et al. 1992; Kellermann et al. 1994). In the radio loud objects, R
generally locates in the range 10 - 1000, and most radio quiet
objects fall in the range 0.1 - 1 (Peterson 1997). An alternative
criterion to distinguish the radio loud and radio quiet objects is
based on the radio luminosity alone, which set a limitation at
$\rm P_{6cm}\approx 10^{25}~ W~ Hz^{-1}~ sr^{-1}$ (Miller, Peacock
\& Mead 1990). According to these criterion, just only $\rm 10\% -
20\%$ was qualified as radio loud objects in optical selected
samples (e.g. Kellermann et al. 1989; Hooper et al. 1995).
However, some studies questioned this distribution by using radio
selected samples (e.g. Wadadekar \& Kembhavi 1999; White et al.
2000). Ho \& Peng (2001) recalculated the radio loudness with the
measurement of nuclear component for a sample of Syfert 1
galaxies, and claimed that at least $\rm 60\%$ of sources in their
sample are characterized by $\rm R\geq 10$, and then territory
traditionally reserved for radio loud AGNs. Similarly, the radio
loudness calculation can also be questioned, due to the fact that
the observed luminosity at 5 GHz of radio loud AGNs has been
extremely Doppler boosted, and the intrinsic luminosity can be
(much) lower. The Doppler enhancement of relativistic flows in
jets is a crucial parameter since the total luminosity of an
unresolved jet scales as the Doppler factor to the fourth power
and to the third power for a resolved cylindrical jet (Lind \&
Blandford 1985). Moreover, the Doppler enhancement of the
synchrotron emission of the jet may also hold at optical bands,
thus the optical emission may be dominated by the beamed
non-thermal synchrotron emission for radio loud AGNs in general,
Flat Spectrum Radio Quasars (FSRQs) in particular. Consequently,
this effect must be considered when the radio loudness is used to
indicate the radio properties for radio loud AGNs, although the
accurate correction for individual object is rather difficult.

The primary goal of this paper is to investigate the issue of
whether the jet power and black hole mass are related in the radio
loud AGNs, by calculating the jet power and black hole mass. We
also try to define a new radio loudness by eliminating the Doppler
boosting effects. Then, we explore the connection between the
radio loudness and black hole mass. $\S$ 2 presents the sample and
$\S$ 3 depicts the methods to estimate jet power, black hole mass,
broad line region luminosity, and new-defined radio loudness. The
results and discussions are shown in $\S$ 4. In last section, $\S$
5, we will give the conclusions. Throughout the paper, we adopt
the spectral index convention $\rm f_{\nu}\propto\nu^{-\alpha}$
and a cosmology with $\rm H_{0}=70 \rm {~km ~s^ {-1}~Mpc^{-1}}$,
$\rm \Omega_{M}=0.3$, $\rm \Omega_{\Lambda} = 0.7$. All values of
luminosity used in this paper are corrected to our adopted
cosmological parameters.

\section{Sample Selection}
We started this work with the quasars and BL Lac objects in the 1
Jy, S4 and S5 radio source catalogues (all sources identified as
galaxies have not been considered here) which has the VLA extended
radio 5 GHz observations. To estimate the black hole mass and
broad line region luminosity, we then searched the literature for
available measurements of the full width at half maximum (FWHM) of
one of broad $\rm H\beta$, MgII, CIV lines, as well as the line flux of
these lines. Finally, the sample of 146 radio loud quasars is
constructed, of which 79 are FSRQs with $\rm \alpha_{2GHz-8GHz}<
0.5$ and 67 are Steep Spectrum Radio Quasars (SSRQs) with $\rm
\alpha_{2GHz-8GHz}> 0.5$.

Table 1 gives the list of our sample with the relevant information
for each object. Columns (1) - (3) represent the object's IAU
name, redshift, and radio spectrum type labeled from the spectral
index between 2 GHz and 8 GHz (FS is for flat spectrum and SS
denotes steep spectrum), respectively. In Columns (4) and (5), we
list the jet power derived from the extended radio flux density
and the references for radio extended flux density, respectively.
Columns (6) - (8) list the broad line region luminosity, the
adopted broad line and its corresponding references, respectively.
In Columns (9) - (11), we present the black hole mass, the adopted
line for FWHM and its references, respectively. The conventional
and new-defined radio loudness are listed in Columns (12) and
(13), respectively (see details in $\S$ 3.3).
A more detailed notation is given at the end of Table 1.

The extended radio luminosity has been K-corrected to 5 GHz in the
source rest frame, assuming $\rm \alpha=1.0$. Fig. 1 shows the
redshift distribution along with the radio total luminosity at 5
GHz (Fig. 1a) and radio extended luminosity at 5 GHz (Fig. 1b). It
can be seen that almost all sources locate in the redshift range
$\rm 0.1<z<2.5$, and both the total and extended 5 GHz radio
luminosities cover about four orders of magnitude. When FSRQs and
SSRQs are nearly indistinguishable in total 5 GHz luminosity, the
extended luminosities of FSRQs are statistically lower than those
of SSRQs.


\begin{figure}
\centerline{\epsfxsize=160mm\epsfbox{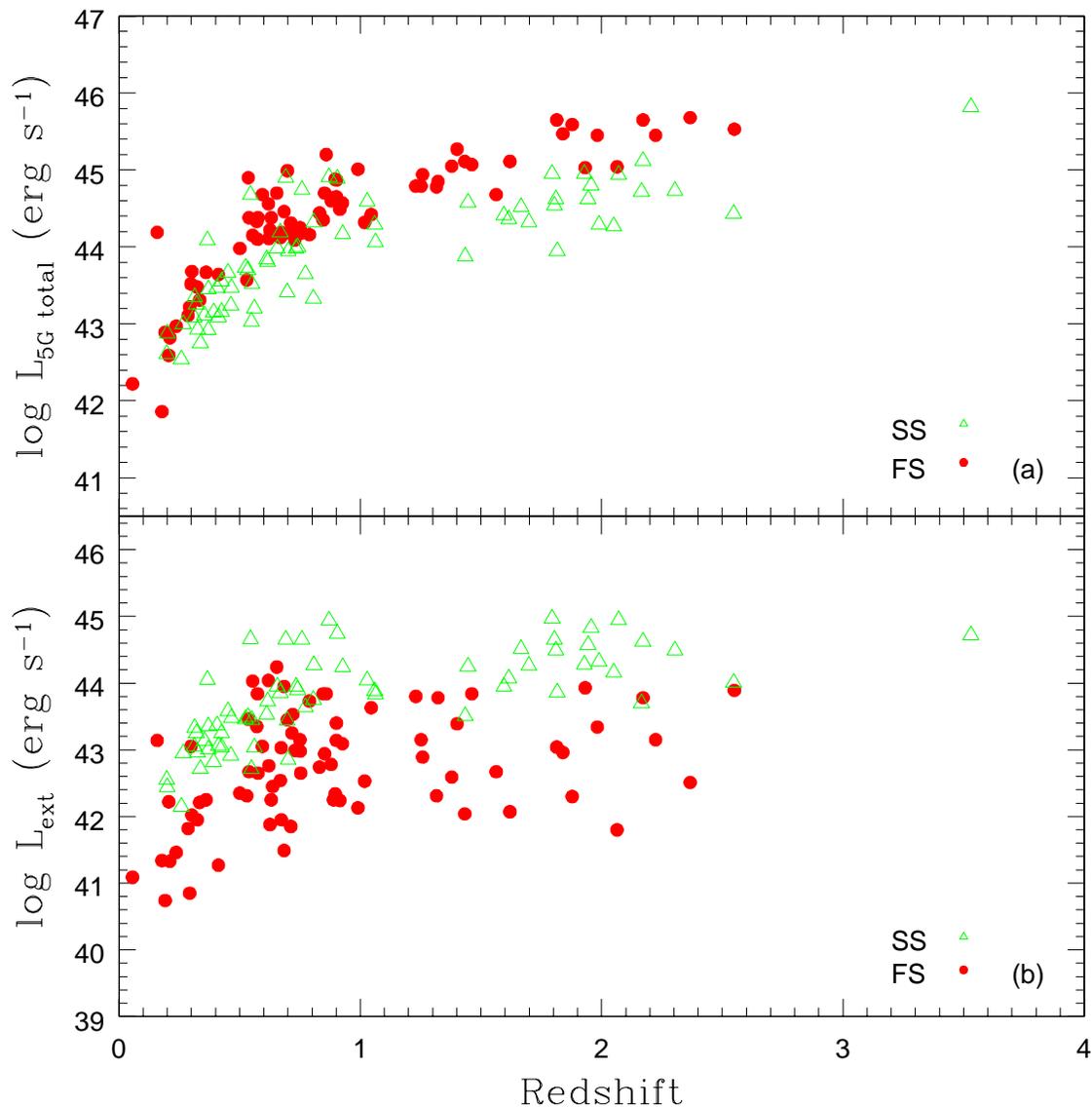}} \caption{(a) The
total 5 GHz radio luminosity versus redshift; (b) The radio extended
luminosity at 5 GHz versus redshift. The solid circles display FSRQs
( $\rm \alpha<0.5$ ), and the open triangles show SSRQs ( $\rm
\alpha>0.5$ ). The spectral index $\alpha$ is derived from 8 GHz to
2 GHz.}
\end{figure}

\clearpage
\begin{deluxetable}{cccccccccccccc}
\tabletypesize{\scriptsize} \tablecaption{The Sample}
\tablewidth{0pt} \tablehead{ \colhead{Object} & \colhead{$z$}
&\colhead{type} &\colhead{log $\rm Q_{jet}$} &\colhead{Refs.}
&\colhead{log $\rm L_{BLR}$}&\colhead{Line} &\colhead{Refs.}
&\colhead {log $\rm
M_{BH}$}&\colhead{Line}&\colhead{Refs.}&\colhead{log R}
&\colhead{log $\rm R_{*}$}} \startdata
0017$+$154& 2.070& SS& 47.19 & BM87& 45.79 &\civ   &   C91 &  9.62 &\mgii &   H02& 3.34 & 3.35 \\
0022$-$297& 0.406& SS& 45.61 &  K98& 44.15 &\hb    &   S93 &  7.81 &\hb   &   G01& 4.40 & 4.30 \\
0056$-$001& 0.717& FS& 45.49 & BM87& 44.91 &\hb    &   B96 &  8.37 &\hb   &   B96& 3.58&  2.85 \\
0101$-$025& 2.050& SS& 46.40 &  R99& 45.69 &\mgii  &   C91 &  8.47 &\mgii &   C91& 3.00&  2.90 \\
0119$+$041& 0.637& FS& 44.69 & BM87& 44.63 &\hb    &  JB91 &  8.65 &\hb   &   G01& 4.34&  2.62 \\
0119$-$046& 1.928& SS& 46.52 &  R99& 46.00 &\mgii  &  SS91 &  9.91 &\mgii &  B94b& 2.92&  2.25 \\
0122$-$042& 0.561& SS& 45.28 &  R99& 44.16 &\mgii  &   W86 &  8.61 &\mgii &   W86& 2.53&  2.37 \\
0133$+$207& 0.425& SS& 45.50 & BM87& 45.02 &\hb    &  JB91 &  9.45 &\hb   &   H02& 3.56&  3.26 \\
0134$+$329& 0.367& SS& 46.30 & BM87& 44.96 &\hb    &  JB91 &  8.65 &\hb   &   C97& 3.53&  3.49 \\
0135$-$247& 0.831& FS& 44.98 & BM87& 45.36 &\hb    &  JB91 &  9.11 &\hb   &   G01& 3.34&  1.87 \\
0159$-$117& 0.669& FS& 44.79 & WB86& 45.34 &\hb    &   O84 &  9.26 &\hb   &   B96& 3.03&  1.90 \\
0212$+$735& 2.367& FS& 44.75 & NH90& 44.95 &\mgii  &   L96 &  6.96 &\mgii &   L96& 4.50&  1.81 \\
0226$-$038& 2.064& FS& 44.03 &  B00& 45.90 &\civ   &   O94 &  8.92 &\hb   &   M99& 2.96&  0.60 \\
0237$-$233& 2.224& FS& 45.38 & BM87& 46.79 &\hb    &  B94a &  9.57 &\hb   &   G01& 3.18&  0.97 \\
0238$+$100& 1.816& SS& 46.10 &  N89& 45.59 &\civ   &   C91 &  9.48 &\civ  &   C91& 2.14&  2.06 \\
0248$+$430& 1.316& FS& 44.55 & CJ01& 45.27 &\mgii  &   S93 &  8.49 &\mgii &  B94b& 3.25&  1.61 \\
0333$+$321& 1.258& FS& 45.13 & BM87& 45.93 &\mgii  &  SS91 &  9.25 &\mgii &  B94b& 2.37&  1.41 \\
0336$-$019& 0.852& FS& 45.18 & BM87& 45.00 &\hb    &  JB91 &  8.89 &\hb   &   G01& 4.06&  2.48 \\
0349$-$146& 0.616& SS& 45.97 & WB86& 45.67 &\hb    &   M96 &  9.97 &\hb   &   M96& 2.64&  2.56 \\
0352$+$123& 1.616& SS& 46.30 &  N89& 45.45 &\mgii  &   C91 &  8.49 &\mgii &   C91& 3.20&  2.90 \\
0403$-$132& 0.571& FS& 45.60 & BM87& 45.25 &\hb    &   O84 &  9.08 &\hb   &   M96& 3.67&  3.02 \\
0405$-$123& 0.574& FS& 46.08 & BM87& 45.91 &\hb    &   O84 &  9.35 &\hb   &   M96& 2.47&  2.51 \\
0406$-$127& 1.563& FS& 44.91 &  N89& 44.96 &\civ   &   W86 &  9.21 &\civ  &   W86& 3.51&  2.32 \\
0413$-$210& 0.808& SS& 46.51 &  K98& 43.99 &\mgii  &   W86 &  8.18 &\mgii &   W86& 3.85&  3.80 \\
0414$-$060& 0.773& SS& 45.88 & BM87& 45.72 &\civ   &   O94 &  9.95 &\hb   &   B96& 2.32&  2.31 \\
0420$-$014& 0.915& FS& 44.48 & BM87& 44.92 &\mgii  &   B89 &  8.41 &\hb   &   B96& 3.43&  1.88 \\
0424$-$131& 2.165& SS& 45.94 & BM87& 46.11 &\civ   &   O94 &  9.76 &\hb   &   M99& 2.88&  1.86 \\
0454$-$220& 0.534& SS& 45.73 & WB86& 45.84 &\hb    &   M96 &  9.57 &\hb   &   M96& 2.71&  2.49 \\
0518$+$165& 0.759& SS& 46.89 & BM87& 45.04 &\hb    &  JB91 &  8.60 &\hb   &   G01& 4.04&  3.95 \\
0537$-$441& 0.896& FS& 44.58 & BM87& 45.05 &\mgii  &   W86 &  8.33 &\mgii &   W86& 3.10&  1.85 \\
0538$+$498& 0.545& SS& 46.90 & BM87& 44.54 &\hb    &   L96 &  9.23 &\hb   &   G01& 3.97&  3.94 \\
0602$-$319& 0.452& SS& 45.82 &  U81& 44.49 &\hb    &   R84 &  9.02 &\hb   &   G01& 3.74&  3.64 \\
0607$-$157& 0.324& FS& 44.20 & BM87& 43.56 &\hb    &   H78 &  7.32 &\hb   &   G01& 3.00&  3.37 \\
0637$-$752& 0.654& FS& 46.48 & CJ01& 45.44 &\hb    &   T93 &  8.81 &\hb   &   G01& 3.26&  2.86 \\
0711$+$356& 1.620& FS& 44.31 &  M93& 45.80 &\mgii  &   L96 &  8.14 &\mgii &   L96& 3.08&  1.02 \\
0723$+$679& 0.846& FS& 46.08 &  U81& 44.80 &\hb    &   L96 &  8.46 &\hb   &   G01& 3.63&  3.69 \\
0736$+$017& 0.191& FS& 42.99 & BM87& 44.18 &\hb    &  JB91 &  7.86 &\hb   &   B96& 2.95&  1.21 \\
0738$+$313& 0.631& FS& 44.50 & BM87& 45.78 &\hb    &  JB91 &  9.57 &\hb   &   B96& 3.20&  1.50 \\
0740$+$380& 1.063& SS& 46.06 & BM87& 45.71 &\civ   &   W95 &  9.09 &\hb   &   H03& 2.92&  2.68 \\
0748$+$126& 0.889& FS& 44.48 &  M93& 44.95 &\mgii  &   W86 &  8.15 &\mgii &   W86& 3.80&  1.86 \\
0802$+$103& 1.956& SS& 47.06 &  A94& 46.24 &\civ   &   C91 &  8.98 &\civ  &   C91& 3.49&  3.52 \\
0804$+$499& 1.433& FS& 44.28 &  M93& 45.39 &\mgii  &   L96 &  9.39 &\mgii &   L96& 3.60&  1.38 \\
0809$+$483& 0.871& SS& 47.18 & BM87& 44.84 &\hb    &   L96 &  8.95 &\hb   &   L96& 4.06&  4.09 \\
0836$+$710& 2.172& FS& 46.01 &  M93& 46.43 &\mgii  &   L96 &  9.36 &\hb   &   M99& 3.34&  2.46 \\
0837$-$120& 0.198& SS& 44.79 &  R99& 45.00 &\hb    &   B96 &  8.86 &\hb   &   B96& 2.48&  2.41 \\
0838$+$133& 0.684& FS& 46.19 & BM87& 45.14 &\hb    &  JB91 &  8.67 &\hb   &   B96& 3.64&  3.57 \\
0850$+$581& 1.322& FS& 46.02 & CJ01& 45.66 &\mgii  &   L96 &  8.49 &\mgii &   L96& 3.63&  2.69 \\
0859$+$470& 1.462& FS& 46.08 &  M93& 45.27 &\mgii  &   L96 &  7.67 &\mgii &   L96& 4.07&  3.14 \\
0903$+$169& 0.412& SS& 45.30 & BM87& 44.69 &\hb    &   B96 &  8.39 &\hb   &   B96& 3.25&  3.21 \\
0906$+$015& 1.018& FS& 44.77 & BM87& 45.11 &\mgii  &   B89 &  8.55 &\mgii &   W86& 3.34&  1.99 \\
0906$+$430& 0.668& SS& 46.09 & BM87& 43.34 &\hb    &   L96 &  6.85 &\hb   &   G01& 3.95&  3.61 \\
0923$+$392& 0.698& FS& 45.69 & BM87& 45.79 &\hb    &   L96 &  9.09 &\hb   &   L96& 4.48&  2.72 \\
0945$+$408& 1.252& FS& 45.39 &  M93& 45.59 &\mgii  &   L96 &  8.60 &\mgii &   L96& 3.50&  2.10 \\
0953$+$254& 0.712& FS& 44.09 & BM87& 44.97 &\hb    &  JB91 &  8.70 &\hb   &   G01& 3.56&  1.43 \\
0954$+$556& 0.901& FS& 45.64 &  M93& 44.98 &\hb    &   L96 &  7.87 &\hb   &   G01& 3.81&  3.62 \\
0955$+$326& 0.530& FS& 44.55 & WB86& 45.35 &\hb    &   M96 &  9.29 &\hb   &   M96& 2.51&  1.69 \\
1007$+$417& 0.612& SS& 45.77 & WB86& 45.71 &\hb    &  JB91 &  9.03 &\hb   &   B96& 2.79&  2.48 \\
1011$-$282& 0.258& SS& 44.39 &  K98& 45.05 &\hb    &   C97 &  8.50 &\hb   &   B96& 2.56&  2.17 \\
1023$+$067& 1.699& SS& 46.50 & BM87& 45.07 &\mgii  &   C91 &  8.99 &\mgii &   C91& 3.10&  3.04 \\
1028$+$313& 0.178& FS& 43.59 & BM87& 44.37 &\hb    &   B96 &  8.44 &\hb   &   B96& 2.40&  1.59 \\
1040$+$123& 1.029& SS& 46.27 & BM87& 45.11 &\mgii  &   N79 &  8.76 &\mgii &   H02& 3.41&  2.86 \\
1100$+$772& 0.312& SS& 45.30 & BM87& 44.83 &\hb    &  JB91 &  9.19 &\hb   &   B96& 2.51&  2.47 \\
1103$-$006& 0.426& SS& 45.28 &  R99& 45.07 &\hb    &   B96 &  9.27 &\hb   &   B96& 2.54&  2.42 \\
1111$+$408& 0.734& SS& 46.19 & BM87& 45.57 &\hb    &  JB91 &  9.84 &\hb   &   G01& 3.45&  3.42 \\
1117$-$248& 0.466& SS& 45.72 &  K98& 43.59 &\mgii  &   W86 &  8.93 &\mgii &   W86& 3.00&  3.00 \\
1136$-$135& 0.554& FS& 46.27 & WB86& 45.20 &\hb    &   O84 &  8.45 &\hb   &   O02& 3.11&  3.73 \\
1137$+$660& 0.656& SS& 46.19 & BM87& 45.85 &\hb    &  JB91 &  9.31 &\hb   &   H02& 2.91&  2.88 \\
1148$-$001& 1.982& FS& 45.57 & CJ01& 46.41 &\civ   &   B89 &  8.90 &\mgii &  B94b& 3.55&  1.87 \\
1150$+$497& 0.334& FS& 44.46 & CJ01& 44.39 &\hb    &   B96 &  8.45 &\hb   &   B96& 3.23&  2.39 \\
1156$+$295& 0.729& FS& 45.23 &  M93& 44.90 &\hb    &   B96 &  8.54 &\hb   &   B96& 3.17&  2.70 \\
1202$-$262& 0.789& FS& 45.97 &  K98& 44.07 &\hb    &   B99 &  8.59 &\hb   &   G01& 4.04&  4.26 \\
1226$+$023& 0.158& FS& 45.38 & BM87& 45.59 &\hb    &  JB91 &  8.92 &\hb   &   C97& 3.14&  2.17 \\
1226$+$105& 2.304& SS& 46.72 &  G91& 46.39 &\civ   &   C91 &  9.65 &\civ  &   C91& 3.17&  2.92 \\
1229$-$021& 1.045& FS& 45.87 & CJ01& 45.68 &\mgii  &   B89 &  8.70 &\mgii &   W86& 3.34&  2.52 \\
1232$-$249& 0.355& SS& 45.38 &  K98& 43.53 &\mgii  &   W86 &  8.91 &\mgii &   W86& 2.95&  2.97 \\
1237$-$101& 0.753& FS& 44.89 & BM87& 44.97 &\mgii  &   S93 &  8.95 &\hb   &   O02& 3.59&  2.24 \\
1250$+$568& 0.321& SS& 45.50 & BM87& 44.57 &\hb    &  JB91 &  8.31 &\hb   &   B96& 3.56&  3.56 \\
1253$-$055& 0.536& FS& 45.70 & BM87& 44.64 &\hb    &   M96 &  8.28 &\hb   &   G01& 4.60&  3.45 \\
1258$+$404& 1.666& SS& 46.75 & CJ01& 45.79 &\mgii  &   C91 &  8.22 &\mgii &   C91& 3.71&  3.70 \\
1302$-$102& 0.286& FS& 44.07 & CJ01& 44.91 &\hb    &   M96 &  8.51 &\hb   &   O02& 2.28&  2.33 \\
1317$+$520& 1.060& SS& 46.12 &  H83& 45.80 &\mgii  &  SS91 &  9.01 &\mgii &  B94b& 2.97&  2.56 \\
1318$+$113& 2.171& SS& 46.86 &  G91& 45.86 &\civ   &   C91 &  9.32 &\civ  &   C91& 3.88&  3.38 \\
1327$-$214& 0.524& SS& 45.70 &  K98& 45.76 &\hb    &   C97 &  9.28 &\hb   &   C97& 2.79&  2.53 \\
1334$-$127& 0.539& FS& 44.91 & CJ01& 44.18 &\mgii  &   S93 &  7.98 &\mgii &   W86& 3.75&  3.25 \\
1354$+$195& 0.720& FS& 45.77 & WB86& 45.93 &\hb    &   B96 &  9.38 &\hb   &   B96& 2.89&  2.25 \\
1355$-$416& 0.313& SS& 45.58 & CJ01& 45.26 &\hb    &   T93 &  9.65 &\hb   &   C97& 2.73&  2.72 \\
1424$-$118& 0.806& SS& 45.99 &  R99& 45.76 &\mgii  &   S89 &  9.18 &\mgii &   W86& 1.99&  2.41 \\
1434$-$076& 0.697& SS& 45.68 &  R99& 44.48 &\mgii  &   W86 &  8.84 &\mgii &   W86& 2.65&  2.68 \\
1442$+$101& 3.530& SS& 46.95 & BM87& 45.93 &\civ   &   C91 &  9.93 &\hb   &   H03& 3.52&  2.42 \\
1458$+$718& 0.905& SS& 46.98 & BM87& 45.47 &\hb    &   L96 &  8.77 &\hb   &   L96& 3.64&  3.49 \\
1502$+$106& 1.839& FS& 45.19 & BM87& 45.57 &\mgii  &   W86 &  8.74 &\mgii &   W86& 4.15&  1.95 \\
1510$-$089& 0.361& FS& 44.50 & BM87& 44.65 &\hb    &   B96 &  8.20 &\hb   &   B96& 3.17&  2.28 \\
1512$+$370& 0.371& SS& 45.25 & WB86& 44.46 &\hb    &   M96 &  8.77 &\hb   &   C97& 2.10&  2.19 \\
1545$+$210& 0.266& SS& 45.20 & BM87& 44.86 &\civ   &   O94 &  8.90 &\hb   &   C97& 3.00&  2.95 \\
1546$+$027& 0.412& FS& 43.52 &  M93& 44.68 &\mgii  &   B89 &  8.47 &\hb   &   O02& 3.55&  1.16 \\
1559$+$173& 1.944& SS& 46.81 &  S90& 45.66 &\mgii  &   C91 &  9.25 &\mgii &   C91& 2.90&  2.85 \\
1606$+$289& 1.989& SS& 46.56 & BM87& 45.61 &\civ   &   C91 &  9.37 &\civ  &   C91& 3.05&  3.09 \\
1611$+$343& 1.401& FS& 45.63 & BM87& 45.91 &\civ   &   W95 &  9.60 &\hb   &   G01& 3.85&  2.10 \\
1618$+$177& 0.551& SS& 45.69 & BM87& 46.13 &\hb    &   M96 &  9.65 &\hb   &   H02& 2.66&  2.59 \\
1622$+$238& 0.927& SS& 46.48 & BM87& 45.34 &\mgii  &  SS91 &  9.53 &\hb   &   B96& 3.27&  3.33 \\
1624$+$416& 2.550& FS& 46.13 &  P95& 43.97 &\civ   &   L96 &  6.35 &\civ  &   L96& 5.11&  4.26 \\
1629$+$120& 1.795& SS& 47.21 &  S90& 45.63 &\civ   &   C91 &  9.65 &\civ  &   C91& 3.62&  3.65 \\
1633$+$382& 1.814& FS& 45.28 & BM87& 45.84 &\mgii  &   L96 &  8.67 &\mgii &   L96& 4.18&  2.04 \\
1637$+$574& 0.750& FS& 45.39 &  M93& 45.57 &\hb    &   L96 &  9.22 &\hb   &   M96& 3.31&  2.38 \\
1641$+$399& 0.594& FS& 45.30 & BM87& 45.47 &\hb    &   L96 &  9.27 &\hb   &   M96& 3.57&  2.38 \\
1642$+$690& 0.751& FS& 45.22 &  M93& 43.86 &\hb    &   L96 &  6.85 &\hb   &   L96& 4.13&  4.09 \\
1655$+$077& 0.621& FS& 45.00 &  M93& 43.62 &\mgii  &   W86 &  7.28 &\mgii &   W86& 4.60&  3.70 \\
1656$+$053& 0.879& FS& 45.02 &  M93& 46.26 &\hb    &   B96 &  9.74 &\hb   &   B96& 3.03&  1.14 \\
1704$+$608& 0.371& SS& 45.60 & BM87& 44.91 &\hb    &   M96 &  9.49 &\hb   &   C97& 2.51&  2.41 \\
1721$+$343& 0.206& FS& 44.47 & WB86& 44.63 &\hb    &   B96 &  8.01 &\hb   &   B96& 2.73&  1.69 \\
1725$+$044& 0.293& FS& 43.09 & BM87& 44.08 &\hb    &   R84 &  7.72 &\hb   &   O02& 3.03&  1.33 \\
1739$+$522& 1.379& FS& 44.83 & BM87& 45.16 &\mgii  &   L96 &  9.32 &\mgii &   L96& 4.01&  1.82 \\
1803$+$784& 0.684& FS& 43.73 &  M93& 44.56 &\hb    &   L96 &  7.92 &\hb   &   L96& 3.55&  1.54 \\
1828$+$487& 0.691& SS& 46.89 & BM87& 45.25 &\hb    &   L96 &  9.85 &\hb   &   G01& 3.85&  3.61 \\
1845$+$797& 0.056& FS& 43.36 & WB86& 42.97 &\hb    &   L96 &  7.75 &\hb   &   L96& 2.66&  2.74 \\
1857$+$566& 1.595& SS& 46.19 &  N89& 45.71 &\mgii  &   C91 &  9.11 &\mgii &   C91& 2.73&  2.27 \\
1928$+$738& 0.302& FS& 44.26 &  M93& 45.18 &\hb    &   M96 &  8.35 &\hb   &   L96& 3.27&  1.50 \\
1954$+$513& 1.230& FS& 46.04 &  K90& 45.39 &\mgii  &   L96 &  9.18 &\mgii &   L96& 3.67&  2.91 \\
1954$-$388& 0.626& FS& 44.12 & CJ01& 44.20 &\hb    &   T93 &  7.99 &\hb   &   O02& 3.38&  2.25 \\
2024$-$217& 0.463& SS& 45.15 &  K98& 43.63 &\mgii  &   W86 &  8.28 &\mgii &   W86& 3.52&  3.19 \\
2044$-$168& 1.932& FS& 46.17 &  N89& 46.20 &\civ   &   O94 &  9.42 &\civ  &   W86& 3.14&  2.51 \\
2120$+$168& 1.805& SS& 46.88 & BM87& 45.57 &\civ   &   O94 &  9.68 &\mgii &   H02& 2.91&  3.01 \\
2121$+$053& 1.878& FS& 44.53 & BM87& 45.90 &\mgii  &  SS91 &  8.78 &\mgii &  B94b& 3.77&  0.96 \\
2128$-$123& 0.501& FS& 44.59 & CJ01& 45.42 &\hb    &   T93 &  9.02 &\hb   &   O02& 2.77&  1.61 \\
2135$-$147& 0.200& SS& 44.69 & BM87& 44.71 &\hb    &  JB91 &  9.03 &\hb   &   M96& 2.79&  2.35 \\
2141$+$175& 0.211& FS& 43.58 & BM87& 44.26 &\civ   &   O94 &  7.95 &\hb   &   M96& 2.57&  2.09 \\
2143$-$156& 0.701& SS& 45.09 & BM87& 44.95 &\mgii  &   W86 &  7.68 &\hb   &   O02& 3.04&  1.95 \\
2145$+$067& 0.990& FS& 44.37 &  M93& 45.78 &\mgii  &   N79 &  8.87 &\mgii &  B94b& 3.44&  0.87 \\
2155$-$152& 0.672& FS& 45.27 & CJ01& 43.70 &\hb    &   S89 &  7.14 &\hb   &   G01& 3.51&  4.23 \\
2201$+$315& 0.298& FS& 45.29 & BM87& 45.46 &\hb    &  JB91 &  8.94 &\hb   &   M96& 2.73&  2.40 \\
2203$-$188& 0.619& FS& 46.28 & CJ01& 44.16 &\mgii  &   S89 &  8.19 &\mgii &   W86& 4.40&  4.44 \\
2208$-$137& 0.392& SS& 45.07 & WB86& 43.65 &\mgii  &   W86 &  7.97 &\mgii &   W86& 2.84&  2.51 \\
2216$-$038& 0.901& FS& 45.38 & BM87& 45.79 &\mgii  &   B89 &  9.08 &\hb   &   G01& 3.40&  2.12 \\
2247$+$140& 0.237& FS& 43.71 & CJ01& 43.83 &\hb    &   C97 &  7.91 &\hb   &   C97& 3.12&  2.22 \\
2251$+$113& 0.326& SS& 45.21 & BM87& 44.90 &\civ   &   O94 &  8.93 &\hb   &   M96& 2.34&  2.36 \\
2251$+$158& 0.859& FS& 46.08 & BM87& 45.68 &\hb    &  JB91 &  8.83 &\hb   &   G01& 3.58&  2.60 \\
2255$-$282& 0.926& FS& 45.33 &  K98& 45.84 &\hb    &   B99 &  8.92 &\hb   &   G01& 3.28&  1.66 \\
2302$-$279& 1.435& SS& 45.75 &  R99& 44.41 &\civ   &   W86 &  8.97 &\mgii &   W86& 2.55&  2.18 \\
2310$-$322& 0.337& SS& 44.95 &  R99& 45.26 &\hb    &   C97 &  9.21 &\hb   &   C97& 2.47&  2.43 \\
2311$+$469& 0.741& SS& 46.15 &  S01& 44.94 &\hb    &  SK93 &  9.36 &\hb   &   G01& 3.12&  3.02 \\
2314$-$116& 0.549& SS& 44.95 &  R99& 43.78 &\mgii  &   W86 &  8.90 &\mgii &   W86& 2.59&  2.26 \\
2320$-$312& 2.547& SS& 46.24 &  R99& 46.09 &\civ   &   C91 &  9.47 &\civ  &   C91& 2.97&  2.55 \\
2335$-$181& 1.446& SS& 46.48 &  R99& 44.61 &\civ   &   W86 &  9.30 &\civ  &   W86& 3.05&  2.71 \\
2344$+$092& 0.673& FS& 44.19 & BM87& 45.50 &\hb    &  JB91 &  8.89 &\hb   &   B96& 2.81&  0.87 \\
2345$-$167& 0.576& FS& 44.90 & BM87& 44.38 &\hb    &  JB91 &  8.47 &\hb   &   G01& 3.90&  2.85 \\
2354$+$144& 1.810& SS& 46.73 &  H83& 45.75 &\hb    &   C91 &  9.37 &\hb   &   C91& 3.17&  3.04 \\

\enddata

Notes: Column (1): IAU source name. Column (2): redshift. Column
(3): spectrum type. Column (4): jet power $\rm Q_{jet}$ in units
of $\rm erg~s^{-1}$. Column (5): references of radio extended flux
density in calculating jet power. Column (6): broad line region
luminosity in units of $\rm erg~s^{-1}$. Column (7): the adopted
lines in calculating broad line region luminosity. Column (8):
references for lines. Column (9): black hole mass in units of $\rm
M_{\odot}$. Column (10): lines for estimating black hole mass.
Column (11): references for lines. Column (12): conventionally
defined radio loudness. Column (13): new-defined radio loudness,
i.e. ratio of the extended radio luminosity to the thermal optical
luminosity
estimated from the broad line luminosity.\\

\vskip 1mm References: A94: Akujor et al. (1994). B94a: Baker et
al. (1994). B94b: Brotherton et al. (1994). B89: Baldwin et al.
(1989). B96: Brotherton (1996). B99: Baker et al. (1999). B00:
Barthel et al. (2000). BM87: Browne \& Murphy (1987). C91: Corbin
(1991). C97: Corbin (1997). CJ01: Cao \& Jiang (2001). G91:
Garrington et al. (1991). G01: Gu et al. (2001). H78: Hunstead et
al. (1978). H83: Hintzen et al. (1983). H02: Hough et al. (2002).
H03: Hirst et al. (2003). JB91: Jackson \& Browne (1991). K90:
Kollgaard et al. (1990). K98: Kapahi et al. (1998). L96: Lawrence
et al. (1996). M93: Murphy et al. (1993). M96: Marziani et al.
(1996). M99: McIntosh et al. (1999). N79: Neugebauer et al.
(1979). N89: Neff et al. (1989). NH90: Neff \& Hutchings (1990).
O84: Oke et al. (1984). O94: Osmer et al. (1994). O02: Oshlack et
al. (2002). P95: Punsly (1995). R84: Rudy (1984). R99: Reid et al.
(1999). S89: Stickel et al. (1989). S90: Saikia et al. (1990).
S93: Stickel et al. (1993). S01: Saikia et al. (2001). SK93:
Stickel \& Kuhr (1993). SS91: Steidel \& Sargent (1991). T93:
Tadhunter et al. (1993). U81: Ulvestad et al. (1981). W86: Wilkes
(1986). W95: Wills et al. (1995). WB86: Wills \& Browne (1986).
\end{deluxetable}

\clearpage

\section{Method}
\subsection{Jet Power}
In this work, we will use the
formula derived from Punsly (2005):
\begin{equation}
\rm{Q_{jet}=5.7\times10^{44}(1+z)^{1+\alpha}Z^{2}F_{151}} ~~
\rm{erg~s^{-1}}
\end{equation}
\begin{equation}
\rm
Z\approx3.31-3.65\times{[(1+z)^{4}-0.203(1+z)^{3}+0.749(1+z)^{2}+0.444(1+z)+0.205]^{-0.125}}
\end{equation}
to estimate the jet power, where $\rm F_{151}$ is the optically
thin flux density from the lobes measured at 151 MHz in units of
Janskys, and the value of $\rm \alpha \approx 1$ is suggested by
the observations (Kellermann, PaulinyToth \& Williams 1969) as a
good fiducial value (see Punsly 2005 for more details).

It should be noted that the equation (1) is only valid when the
optically thin extended emission is measured. In the radio
selected sample, the radio emission at low frequency, e.g. 151
MHz, is usually dominated in the steep spectrum quasars. That is
to say, the spectrum at low frequency is steep, which usually
emerges from the non-beamed optically thin lobes. In actuality,
most of our objects indeed show a steep spectrum at low frequency.
For these sources, roughly, the radio 151 MHz flux density could
be directly used in equation (1). However, we also find the flat
spectrum at low frequency in some objects. This implies that the
radio emission at low frequency is mainly optically thick, then
the Doppler boosting still holds even at low frequency in these
objects, which precludes the possibility of directly using the
measured flux density in equation (1).

The extended radio flux measured from the optically thin lobes is
free from the Doppler boosting effects, since the lobe material is
generally thought to be of low enough bulk velocity. To further
explore the validity of directly using the measured radio low
frequency flux density in equation (1), we investigate the
relationship between the flux density at 151 MHz and the extended
flux density at 5 GHz in Fig. 2. It is apparent that there is a
tight relationship between them for the sources with steep
spectrum at low frequency (denoted by open triangles and open
circles), since both of them are from the non-beamed optically
thin lobes. However, large scatter and deviation exist in the
sources with flat spectrum at low frequency (shown as solid
triangles and solid circles). This is not surprising since the
extended 5 GHz emission is from radio lobes, whereas the 151 MHz
emission is mainly from radio cores, which is extremely Doppler
boosted. These results show that it is problematic to directly use
the low frequency radio flux density in equation (1), although it
might be valid for the sources with steep spectrum at low
frequency. Due to this fact, we extrapolate the extended 5 GHz
flux density to calculate the extended 151 MHz flux density, by
assuming a spectral index of $\alpha=1.0$, in present work. The
jet power is then estimated by substituting the extrapolated
extended 151 MHz flux density, instead of the measured 151 MHz
flux density, in equation (1). The estimated jet power is listed
in Column (4) of Table 1, where the references of the extended
flux density at 5 GHz are given in Column (5).

\begin{figure}
\centerline{\epsfxsize=160mm\epsfbox{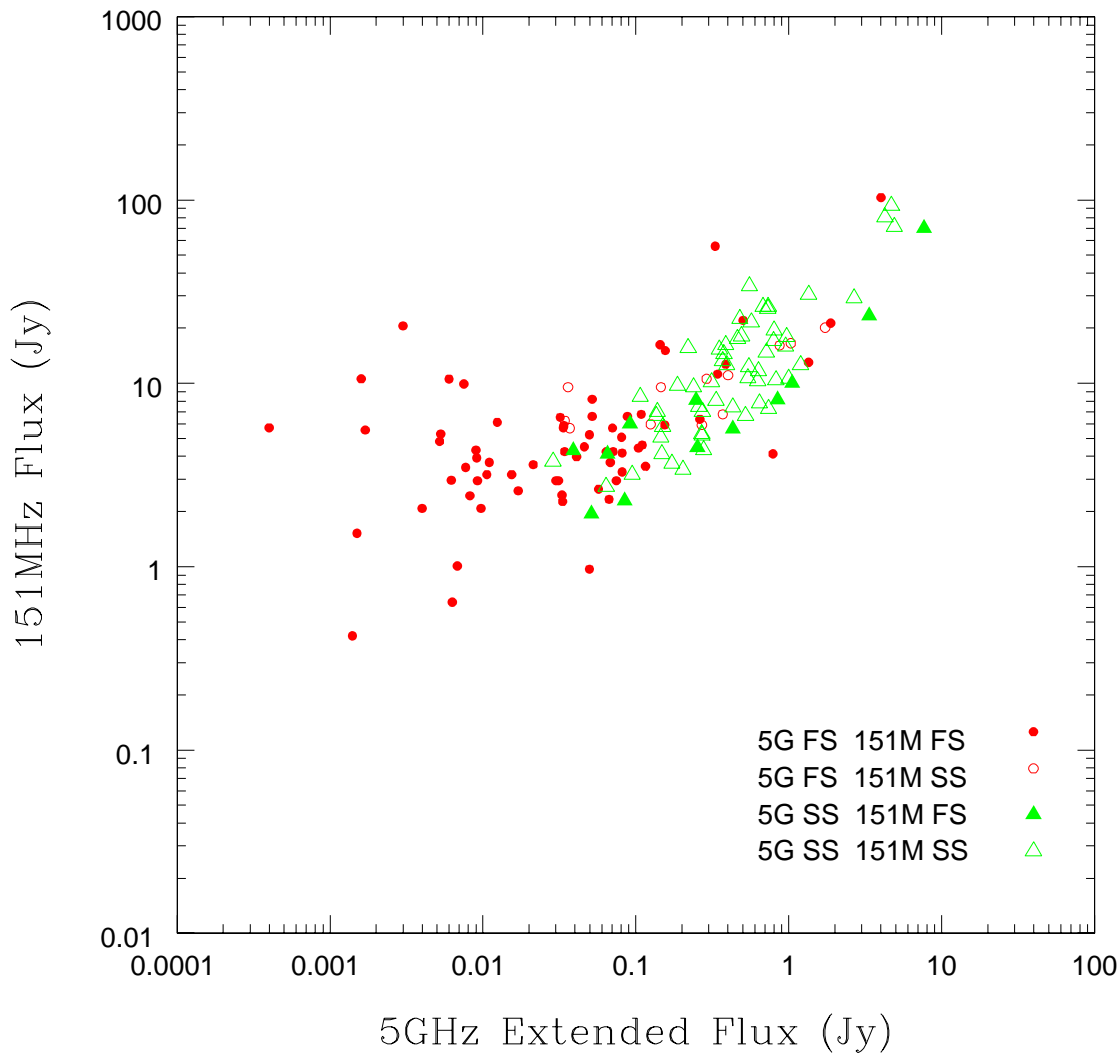}} \caption{The
relationship between the radio 5 GHz extended flux and the radio low
frequency 151 MHz flux. The open triangles display the sources with
steep spectrum both at high frequency (5 GHz) and low frequency (151
MHz); the solid triangles are the sources with steep spectrum at
high frequency (5 GHz) but flat spectrum at low frequency (151 MHz);
the open circles show those with flat spectrum at high frequency (5
GHz) but steep at low frequency (151 MHz); the solid circles denote
those with flat spectrum both at high (5 GHz) and low frequency (151
MHz).}
\end{figure}

\subsection{Black hole mass and Broad line luminosity}

The black hole mass can be estimated on the assumption that the
dominant mechanism responsible for the width of the broad emission
line is the gravitational potential of the central supermassive
black hole, and that the line widths reflect the Keplerian
velocities of the line-emitting material in a virialized system
(Wandel, Peterson \& Malkan 1999; McLure \& Dunlop 2001). The black
hole mass is given by
\begin{equation}
\rm{M_{BH}=R_{BLR}V^2G^{-1}}
\end{equation} where G is the
gravitational constant, V is the velocity of the gas of broad
emission line region gravitationally bound to the central black
hole, and $\rm R_{BLR}$ is the size of the broad line region
(BLR). V can be obtained directly form FWHM of the broad emission
lines ($\rm V=f\times V_{FWHM}$), where f is the factor that
depends on the geometry and kinematics of the broad line region
(McLure \& Dunlop 2002; Vestergaard 2002). In present work, we
adopt an uniform value $\rm f=\sqrt{3}/2$ assuming an isotropic
distribution of the broad line region clouds (Wandel 1999; Kaspi
et al. 2000; Vestergaard 2002).

Besides estimating V from FWHM of the broad emission line, we need
to estimate $\rm R_{BLR}$. The reverberation mapping technique is
the most reliable method to calculate $\rm R_{BLR}$. The basic
concept of this method is using the time lag of the emission line
light with respect to the continuum light to estimate the light
crossing size of the BLR. However, this method can be only used for
a limited number of objects (Kaspi et al. 2000; Onken \& Peterson
2002), since it requires intensive monitoring of the broad lines and
the continuum.

The alternative and extensively used method to estimate $\rm
R_{BLR}$ is to use the empirical relationship found between $\rm
R_{BLR}$ and the optical continuum luminosity, which is inferred
from the $\rm R_{BLR}$ determined from reverberation mapping
method (Kaspi et al. 2000). Kaspi et al. (2000) have calibrated an
empirical relation between the BLR size and the monochromatic
luminosity at 5100 $\rm \AA$. Recently, Kaspi et al. (2005)
reinvestigated the relationship between the characteristic $\rm
R_{BLR}$ and the optical continuum luminosities, making use of the
best available determinations of $\rm R_{BLR}$ for a large number
of AGNs from Peterson et al. (2004). Simply averaging the
measurements obtained from the BCES and FITEXY methods for the
relation between $\rm R_{BLR}$ and the optical luminosity at $\rm
5100~\AA$, using one data point per object (see Kaspi et al.
2005), they found:
\begin{equation}
\rm{R_{BLR}=22.30~(\frac{\lambda L_{\lambda}(5100
\rm{\AA})}{10^{44}~ erg~s^{-1}})^{0.69} ~~\rm{lt-days}}
\end{equation}
Combining the $\rm R_{BLR}$ estimated from this formula and the
FWHM of broad $\rm H\beta$ line measured from single-epoch
spectrum, we can calculate the black hole mass using equation (3).
However, $\rm H\beta$ line will be redshifted out of optical
domain for the objects with relatively high redshift. Thus, this
method is highly limited for high redshift objects (actually, $\rm
H\beta$ line of six high redshift sources in our sample have been
measured from IR spectrum, but usually IR spectrum is not readily
available).
McLure \& Jarvis (2002) presented the expression to estimate black
hole mass through calibrating the empirical relationship between
the $\rm R_{BLR}$ and optical continuum luminosity for broad MgII.
\begin{equation}
\rm \frac{ M_{BH}} {M_{\odot}}  =3.37\left(\frac{\lambda
L_{3000}}{10^{44}~{\rm
~erg~s^{-1}}}\right)^{0.47}\left(\frac{FWHM(MgII)} {{\rm
km~s}^{-1}}\right)^{2} \label{final}
\end{equation}
This enables us to estimate black hole mass for high redshift
AGNs, using the measured FWHM of broad MgII. Moreover,
reverberation studies indicate that the size of the BLR for C IV
is about half that of $\rm H\beta$ (Stirpe et al. 1994; Korista et
al. 1995; Peterson 1997; Peterson \& Wandel 1999). If we assume
that $\rm R_{BLR}(CIV)=0.5R_{BLR}$, where $\rm R_{BLR}$ is derived
from equation (4) (Corbett et al. 2003; Warner et al. 2003, 2004;
Dietrich \& Hamann 2004), then we can estimate the black hole mass
with the FWHM of broad CIV.

Similar as the radio luminosity, the optical luminosity of radio
loud quasars can also be contaminated by the effect of
relativistic beaming. In fact, the optical emission of radio loud
quasars is a mixture of thermal and non-thermal emission.
In general, SSRQs tend to be orientated
with jets pointing away from our line of sight, and FSRQs tend to
be orientated with their jets beamed along our line of sight,
although not explicitly applicable on a source-by-source basis.
Thus, the optical
emission may be dominated by the beamed non-thermal synchrotron
emission in FSRQs. However, for SSRQs, even if the non-thermal
synchrotron emission can extend to optical band, the steep
spectrum and very weak beaming effect make it hardly possible to
dominate over the thermal emission. Since the relationship between
the BLR radius and optical continuum luminosity is supposed to be
valid in the case of a thermal continuum, we thus must estimate
the thermal optical continuum luminosities. In present work, we
will calculate the thermal optical luminosity only for FSRQs. For
SSRQs, we will use the optical luminosity directly as thermal
optical luminosity, since the thermal optical emission dominates
in these sources.

In the radio quiet AGNs, the optical emission is believed to be
free from the contamination of non-thermal synchrotron emission
since their jets are very weak (if present). The broad emission
line emission, which is produced by the illumination of ionizing
luminosity from central AGNs on the BLR gas, can be used as a good
indicator of the thermal optical emission. Therefore, we can
estimate the thermal optical luminosity for FSRQs, by assuming
that their thermal optical luminosities exhibit a dependence on
broad emission line luminosities similar to that of radio quiet
AGNs. We fitted the $\rm H\beta$ line (in $\rm erg~ s^{-1}$) and
optical luminosities (in $\rm erg~ s^{-1}$) to a power-law using
the Ordinary Least Square (OLS) bisector method on the sample of
radio quiet AGNs of Kaspi et al. (2000), and obtained the
dependence:
\begin{equation}
\rm {L_{5100\rm{\AA}}=0.843\times10^{2}~L_{H\beta}^{0.998}}
\end{equation}
The relation of broad $\rm H\beta$ line and optical continuum
luminosity and the fitted line is shown in Fig. 3. For FSRQs with
the optical continuum luminosity exceeding the power-law
dependence of equation (6), we adopt the continuum luminosities
computed with equation (6) at the corresponding broad $\rm H\beta$
line luminosity.
When $\rm H\beta$ is not available, we firstly calibrate the MgII
or CIV lines to $\rm H\beta$ line adopting the relative flux of
the relevant lines of the composite spectrum of Francis et al.
(1991), respectively, then use equation (6) to estimate the
thermal optical luminosity when the optical continuum luminosity
exceeds the power-law dependence of equation (6). Hereafter, we
will call the thermal optical luminosity estimated in this way as
the thermal corrected optical luminosity. The relationship between
the thermal corrected optical luminosity and measured optical
luminosity for FSRQs is shown in Fig. 4. Due to the significant
beaming effect, the large deviations from the equivalent line is
apparently seen. This strengthens the necessity of removing
beaming effect in optical band for FSRQs.

\begin{figure}
\centerline{\epsfxsize=160mm\epsfbox{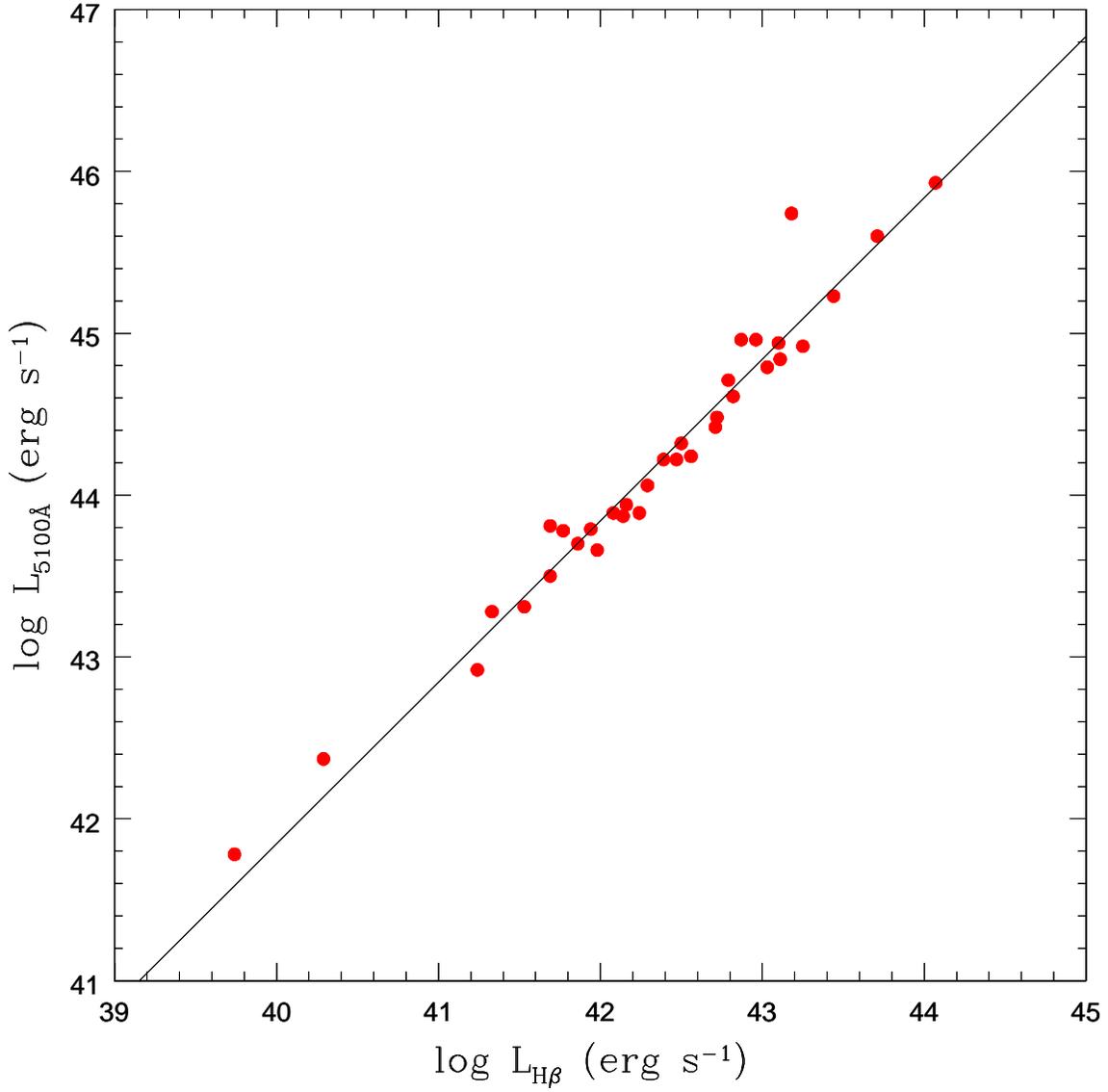}} \caption{The broad
$\rm H\beta$ line luminosity and the optical continuum luminosity at
$\rm 5100~\AA$ for a sample of radio quiet AGNs of Kaspi et al.
(2000). The solid line is the fitted line using OLS method. Data are
from Kaspi et al. (2000), and has been transferred to our adopted
cosmology frame.}
\end{figure}

\begin{figure}
\centerline{\epsfxsize=160mm\epsfbox{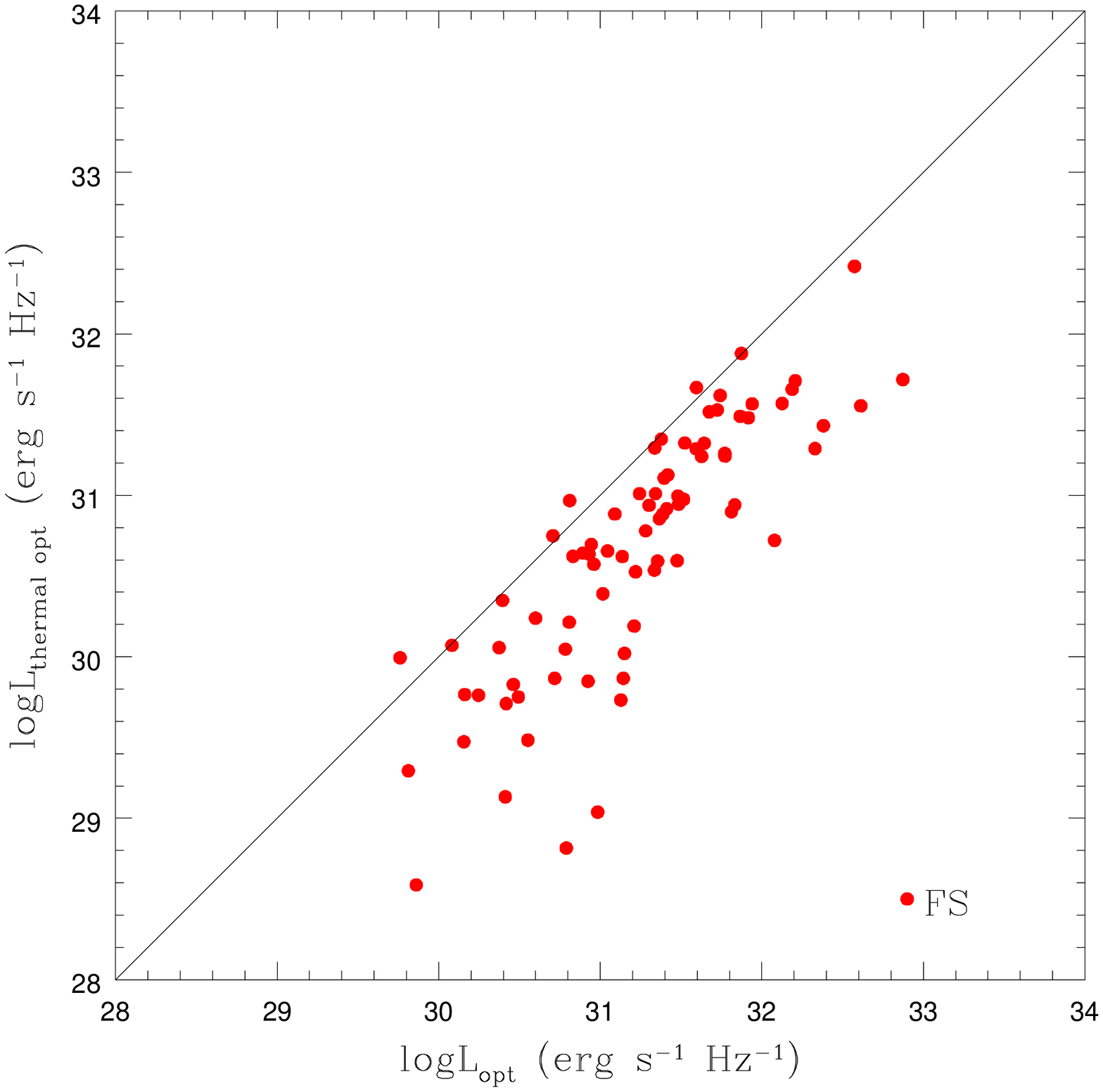}} \caption{The
thermal optical luminosity estimated using equation (6) and
measured optical luminosity for FSRQs. The line is the equivalent
line.}
\end{figure}

After computing the thermal corrected optical continuum luminosity
of FSRQs and the optical luminosity of SSRQs (uncorrected), we
then estimate the black hole mass by using equation (4) and FWHM
of broad $\rm H\beta$ for 92 objects. When only Mg II line is
available, we use FWHM of broad MgII line in combination with
equation (5). For the rest 11 sources, we estimate the BLR size of
CIV assuming that $\rm R_{BLR}(CIV)=0.5R_{BLR}$, then combine FWHM
value in equation (3) to estimate the black hole mass. Although
this assumption may bring uncertainties in black hole mass, the
small number fraction ($\sim7.5\%$) may not influence our results.
The calculated black hole mass, the line adopted and the
corresponding references are shown in Table 1.

Following Celotti et al. (1997), the BLR luminosity is derived by
scaling several strong emission lines to the quasar template
spectrum of Francis et al. (1991), in which $\rm Ly\alpha$ is used
as a reference. From this approach,
\begin{equation}
\rm L_{\rm BLR}=\left\{
\begin{array}{l}
\frac{\rm \langle L_{\rm BLR}\rangle}{\rm L_{\rm est}(\rm H\beta)}=25.26\rm L_{\rm H\beta},\\
  \\
\frac{\rm \langle L_{\rm BLR}\rangle}{\rm L_{\rm est}(\rm Mg II)}=
16.35\rm L_{\rm Mg~ {\small II}},\\
  \\
\frac{\rm \langle L_{\rm BLR}\rangle}{\rm L_{\rm est}(\rm C IV)}=
8.82\rm L_{\rm C~ {\small IV}},
\end{array}
\right.
\end{equation}
where $\rm \langle L_{\rm BLR}\rangle=555.77$, $\rm L_{\rm
est}(\rm H\beta)=22$, $\rm L_{\rm est}(\rm Mg II)=34$, and $\rm
L_{\rm est}(\rm C IV)=63$. Table 1 gives the values of $\rm L_{\rm
BLR}$ estimated from Equation (7) in Column (6), as well as the
adopted line in Column (7).

\subsection{Radio loudness}
The deficit of radio loudness in radio loud quasars lies in the
effect of relativistic beaming both in radio and optical emission,
especially for FSRQs. Actually, it has already been found that
some radio loud quasars can be classified as radio quiet if the
beaming effect of the jets are considered (so called `radio
intermediate quasars', see Falcke, Sherwood \& Patnaik 1996;
Jarvis \& McLure 2002).
To remove the beaming effect, it would be more reasonable to use
the ratio of the intrinsic radio luminosity to thermal optical
luminosity to estimate radio loudness, which can be a good
indicator of the relative importance of radio emission.

In this work, we define a new radio loudness $\rm R_{*}$ as the
ratio of the radio extended luminosity to the thermal corrected
optical continuum luminosity to eliminate the effect of
relativistic beaming. The conventional radio loudness R, and the
new-defined radio loudness $\rm R_{*}$ are shown in Columns (12)
and (13) of Table 1, respectively.
These radio loudness are used to explore the relationship between
them and the black hole mass, and the significance of beaming
effect corrections in $\S$ 4.2.

\section{Results and Discussions}
\subsection{Jet power, BLR luminosity and Black hole mass}

\begin{figure}
\centerline{\epsfxsize=100mm\epsfbox{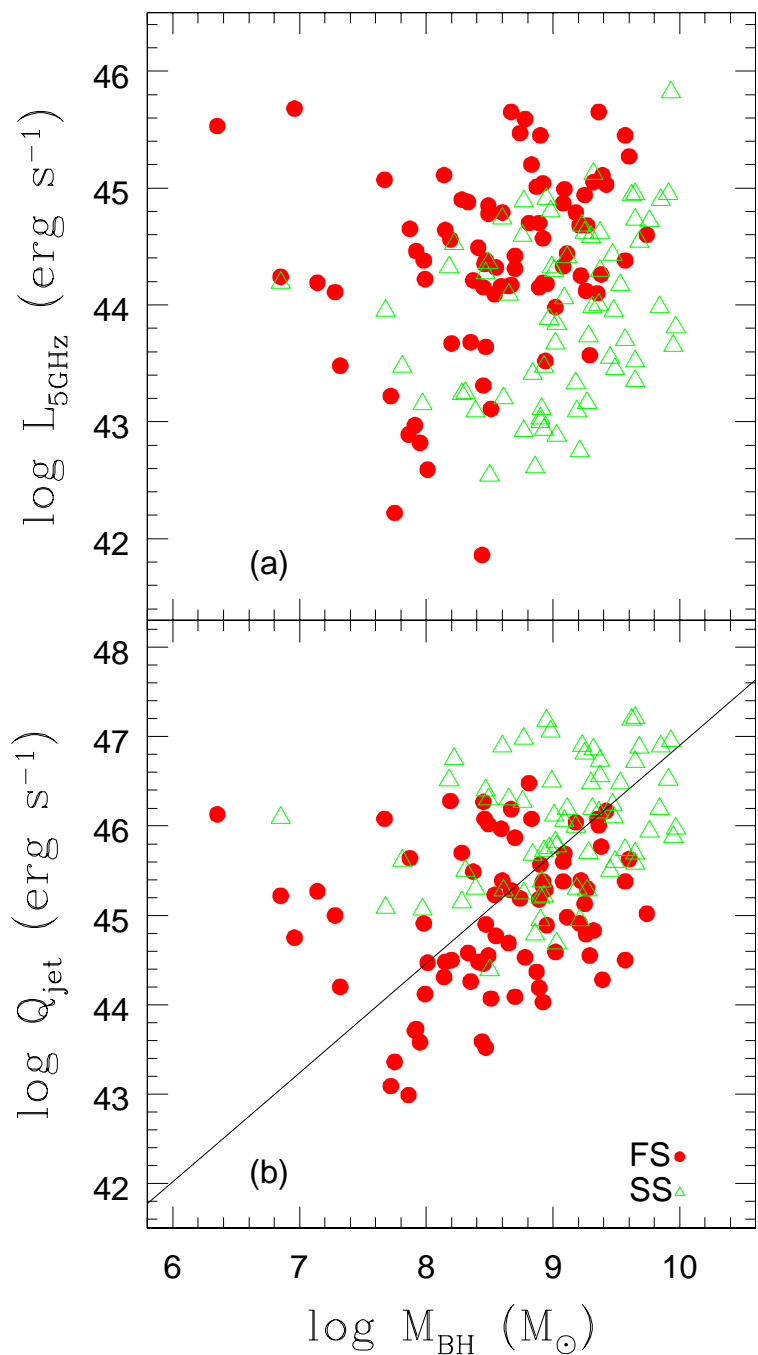}} \caption{(a) The
total radio 5 GHz luminosity vs. black hole mass. (b) The jet
power vs. black hole mass. The solid circles denote FSRQs and the
open triangles display SSRQs. The line is the fitted line using
OLS method.}
\end{figure}

\begin{figure}
\centerline{\epsfxsize=160mm\epsfbox{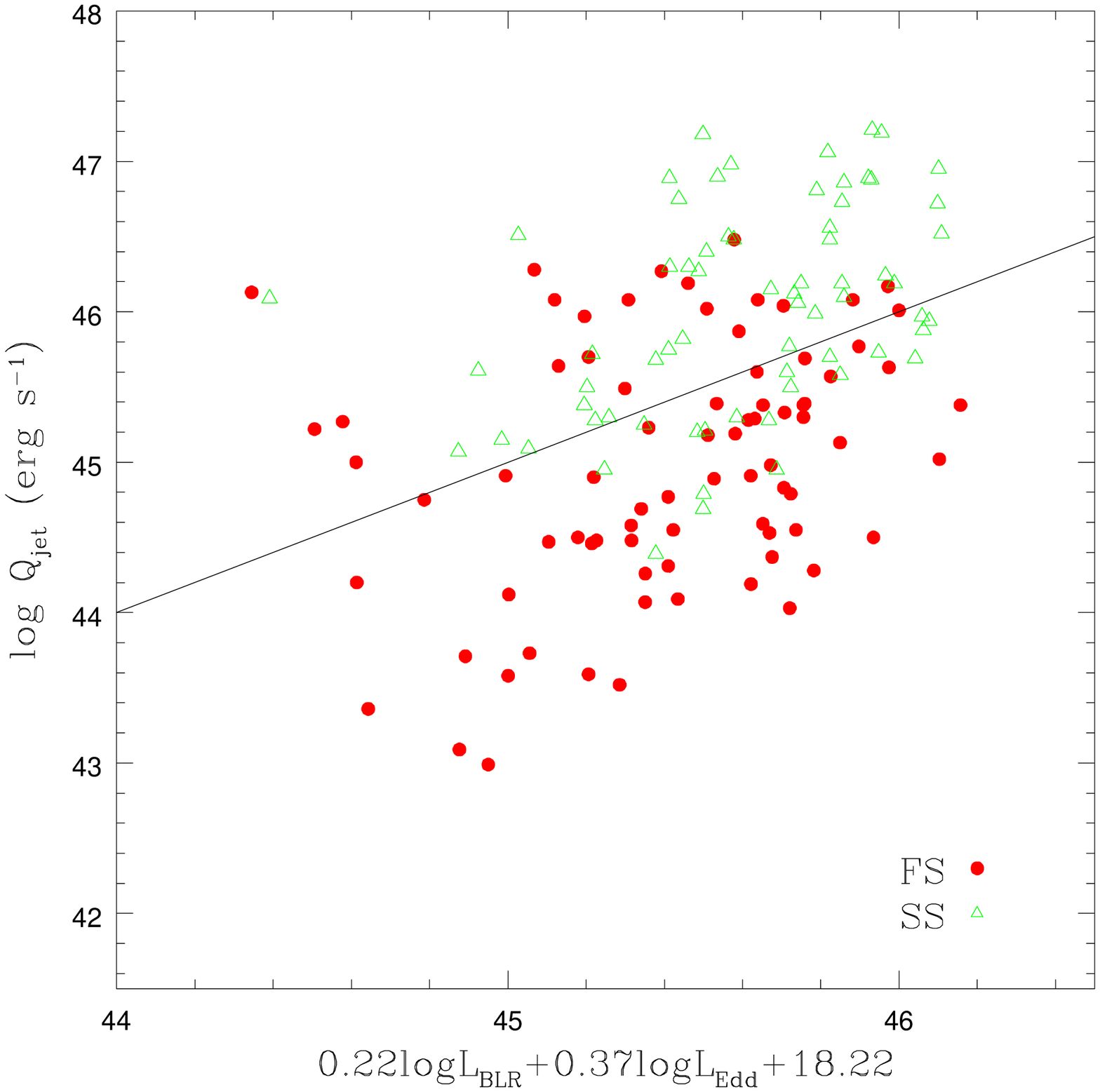}} \caption{The
relationship between the jet power and both the Eddington
luminosity and the broad line region luminosity. The symbols are
the same as in Fig. 5. The line is the least square multivariate
regression line (equation (8)).}
\end{figure}

The relationship between the jet power and black hole mass
is shown in Fig. 5b. The triangles and circles represent the steep
spectrum and flat spectrum quasars, respectively.
We find a significant correlation between the $\rm Q_{jet}$ and
$\rm M_{BH}$ with a correlation coefficient of $\rm r=0.419$ at
$\rm \gg99.99$ per cent confidence.
It should be noted with caution that this correlation may be
caused by the common dependence of redshift. We therefore use the
partial Spearman rank correlation method (Macklin 1982) to check
this correlation. Still, the significant correlation with a
correlation coefficient of 0.336
is present between $\rm Q_{jet}$ and $\rm M_{BH}$,
independent of the redshift. The significance of the partial rank
correlation is 4.167,
 which is equivalent to the
deviation from a unit variance normal distribution if there is no
correlation present. So, the very strong correlation still holds
while subtracting the common dependence of redshift. We conclude
that this correlation is intrinsic for our sample of radio loud
quasars. It therefore strongly supports the scenario of a tight
connection between the relativistic jet and the black hole mass.
The slope of the correlation of $\rm log~Q_{jet}$ against $\rm
log~M_{BH}$ fitted using OLS method is $1.22\pm0.09$.

We also plot the relation of total 5 GHz radio luminosity against
black hole mass in Fig. 5a. The Spearman rank correlation analysis
shows a weak correlation with correlation coefficient $\rm
r=0.173$ at 96.3 per cent confidence. Again, we use the partial
Spearman rank correlation method to check this correlation by
excluding the common dependence of redshift. However, we find no
indication of correlation when subtracting the affecting of
redshift. Thus, the apparent weak correlation might be a
consequence of common dependence of redshift. In view of the
significant intrinsic correlation between the jet power and black
hole mass, it is likely that the different Doppler boosting in
total radio luminosity, as well as the conversion rate of jet
power to intrinsic radio luminosity from source to source,
preclude the correlation to present between the total radio
luminosity and black hole mass.

Motivated by the suggestion of the kinetic luminosity depending on
both the disk luminosity and black hole mass (Wang, Luo \& Ho
2004),
we investigate the correlation between the jet power and both
broad line region luminosity and black hole mass (in terms of its
Eddington luminosity $\rm L_{Edd}$). Using the least-squares
method of multivariate regression, we find that $\rm Q_{jet}$
correlates with both $\rm L_{BLR}$ and $\rm L_{Edd}$ with
correlation coefficient $\rm r=0.432$ and a probability of $\rm
p=8.52\times10^{-8}$ for rejecting the null hypothesis of no
correlation. The relationship can be expressed as:
\begin{equation}
\rm{log~Q_{jet}=(0.22\pm0.13)~log~
L_{BLR}+(0.37\pm0.14)~log~L_{Edd}+(18.22\pm4.99)}
\end{equation}
This correlation (Fig. 6) suggests that the jet power depends on
both the disk luminosity and black hole mass.

By defining the Eddington ratio as $\rm \lambda=L_{bol}/L_{Edd}$,
and assuming $\rm L_{bol}\approx10L_{BLR}$, Equation (8) can be
expressed in a different form as
\begin{equation}
\rm{log~Q_{jet}=0.22~log~\lambda+0.59~log~(\frac{M_{BH}}{M_{\odot}})+40.48}
\end{equation}
This implies that the jet power depends on both the Eddington
ratio and black hole mass, and the black hole mass plays a
dominant role in producing jet power, compared with Eddington
ratio.

Despite the strong correlation exhibited in Fig. 6, clearly there
is significant scatter. A significant fraction of this scatter may
be due to measurement and/or systematic errors in $\rm Q_{jet}$,
$\rm L_{BLR}$, and $\rm M_{BH}$. When estimating the jet power,
the extended 151 MHz flux density has been extrapolated from the
VLA extended 5 GHz flux density, by adopting a spectral index
$\alpha=1.0$. An adoption of $\alpha=0.7$ will bring a factor of
about 3 uncertainty in the jet power. Moreover, it is likely that
the part of low surface brightness radio emission could be
undetectable in VLA radio images for some sources, which can also
bring uncertainties in the jet power. Furthermore, the observed
extended radio emission has been dissipated over a long period,
thus, the jet power estimated from the past radio activity can be
different from that of the time, at which the optical emission are
observed. On the other hand, the uncertainty of black hole mass
estimate using the line width-luminosity-mass relation is
approximately a factor of 3 (Gebhardt et al. 2000; Ferrarese et
al. 2001). However, it seems that all these uncertainties can not
explain the large scatter in Fig. 6. Therefore, it is possible
that the additional physical parameters, such as black hole spin
or source environment, must also be included.

The beaming effect in radio luminosity has been noted by the Jarvis \&
McLure (2002) revisit of the radio luminosity - black hole mass
relation after correcting Doppler boosting on the FSRQs sample of
Oshlack et al. (2002). They found that the FSRQs occupy a wide
range in intrinsic radio luminosity, and that many sources would
be more accurately classified as radio-intermediate or radio-quiet
quasars. Moreover, they claimed that FSRQs are fully consistent
with an upper boundary on radio power of the form $\rm
L_{5GHz}\propto M_{BH}^{2.5}$. However, it should be noted that
the Doppler boosting was only corrected averagely by adopting the
average viewing angle of FSRQs ($7^{\circ}$) and SSRQs
($37^{\circ}$) (see Jarvis \& McLure 2002 for more details). As
the authors argued, both smaller and larger viewing angle are
undoubtedly consistent with a Doppler boosting paradigm (and many
of the sources will have $\theta<7^{\circ}$) for FSRQs.
Recently, Woo et al. (2005) found that the radio
luminosity and black hole mass are not correlated for a sample of
BL Lac objects and FSRQs. Since an accurate beaming correction is
not possible for individual object, the authors simply used the
beam uncorrected radio luminosity. In spite of that, they argued
that any beaming correction would unlikely reveal a hidden
correlation between black hole mass and radio luminosity, given
the fact that radio luminosity between FSRQs and BL Lac objects is
different by a minimum of several orders of magnitude for the
given black hole mass range.
In this paper, we use the radio extended luminosity to indicate
the intrinsic radio luminosity, since it is not affected by
beaming effect. Even so, however, it can't be ignored that the
radio luminosity is merely an indirect measure of the energy
transported through the jets from the central engine, and most of
the energy in the jets is not radiated away but is transported to
the lobes. Therefore, it's essential to explore the relationship
between the fundamental radio parameter, namely, the jet power and
the black hole mass.
Our results in Fig. 5b show a very strong correlation between the
black hole mass and the jet power for a sample of radio loud
quasars. This intrinsic correlation indicates that the jet
formation is closely connected with black hole mass.

By using the kinetic luminosity $\rm L_{Kin}$ calculated by
Celotti et al. (1997), Wang et al. (2004) found $\rm L_{Kin}$
correlates with both $\rm L_{BLR}$ and $\rm M_{BH}$. Moreover,
they argued that the significant correlation between the kinetic
luminosity and broad line region luminosity improved when the
second parameter, $\rm M_{BH}$ is included. We have already shown
that there is a significant correlation between the jet power and
black hole mass, however, this correlation does not improve when
including the broad line region luminosity.
Notwithstanding this, interestingly, we find our results of
equation (8) are consistent with that of Wang et al. (2004) within
$\rm 1\sigma$, in spite of the fact that we estimate the jet power
on the sample of 146 radio loud quasars, whereas the kinetic
luminosity is used in Wang et al. (2004) for a sample of only 35
blazars. The different fundamental radio parameters used on the
different samples leading to the coherent results, suggests that
the relationship between $\rm Q_{jet}$, on both $\rm L_{BLR}$ and
$\rm M_{BH}$ is likely fundamental in radio loud quasars, and the
veracity of these two methods in calculating the power transported
by the radio jets. Nevertheless, this relationship still need to
be confirmed with a larger, and complete sample of radio loud
quasars. In addition, it's more crucial to explore whether such
relationship also holds for radio quiet AGNs, which might help us
understand the difference between two populations.

\subsection{Radio loudness and black hole mass}
\begin{figure}
\centerline{\epsfxsize=160mm\epsfbox{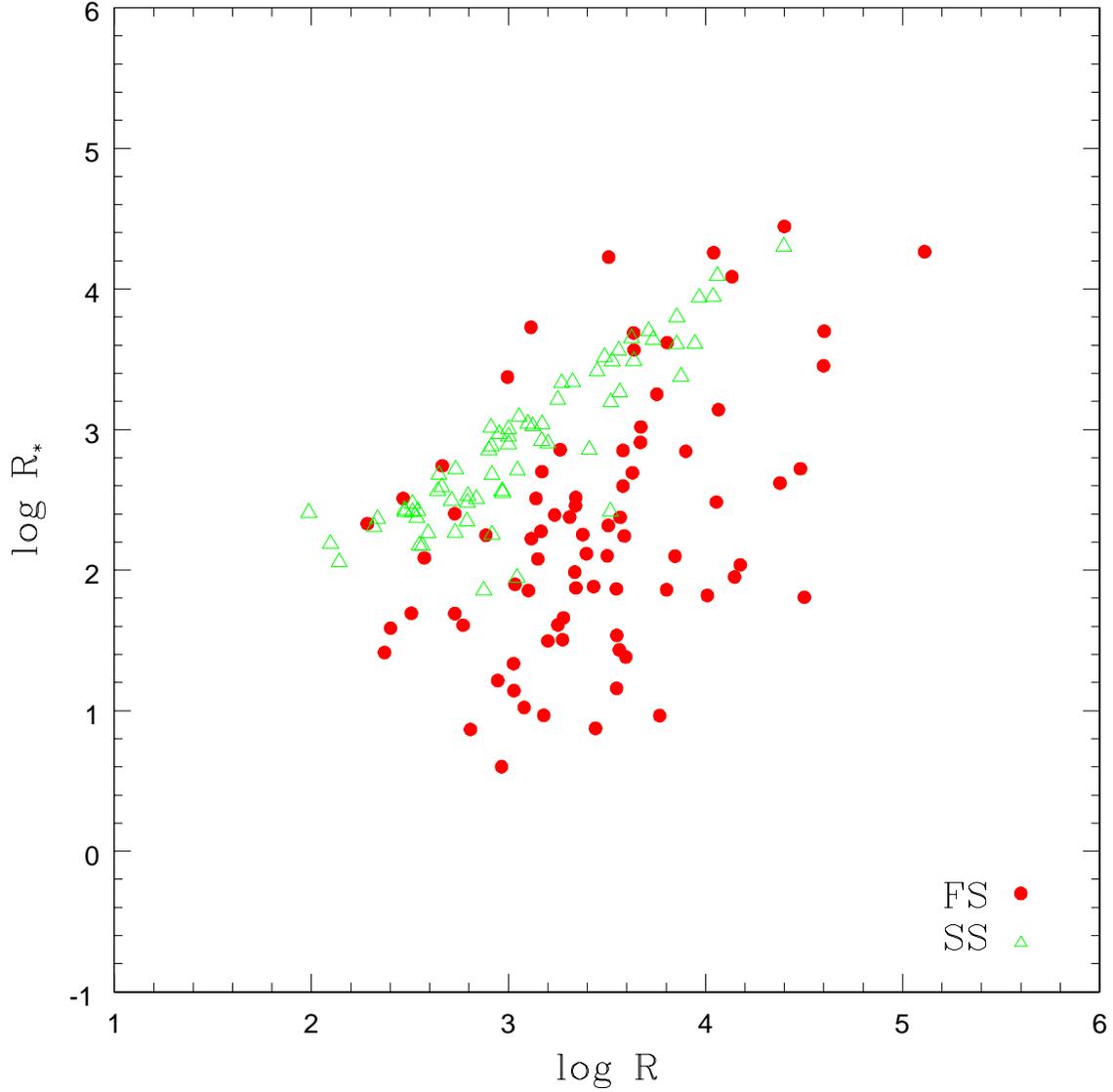}} \caption{The
relationship between R and $\rm R_{*}$. R is the conventionally
defined radio loudness, whereas $\rm R_{*}$ is the new-defined
radio loudness as the ratio of radio extended luminosity to
thermal optical continuum luminosity estimated from the broad line
luminosity (see text). The symbols are the same as in Fig. 5.}
\end{figure}

The comparison between R and $\rm R_{*}$ is shown in Fig. 7. It
should be kept in mind that for new-defined radio loudness, using
the radio extended luminosity to replace the total luminosity
tends to decrease radio loudness, whereas using the thermal
corrected optical luminosity tends to increase. Since corrections
of optical luminosity in SSRQs are not committed presently,
and their extended radio luminosity dominate in the total radio
luminosity, SSRQs will be expected to stay tightly around $\rm
R_{*}=R$. This is apparently seen in Fig. 7. However, some sources
stay somewhat away, but not much. This may be due to the
variations of radio emission, or miss classification.

Conversely, the location of FSRQs in $\rm R-R_{*}$ panel may vary
from source to source, and depends on the amount of corrections in
optical and radio luminosity since both of them are expected to be
significant. This is obviously seen in Fig. 7, where a larger
scatter presents for FSRQs. We find that for almost all FSRQs,
$\rm R_{*}$ are smaller than R, which means that the beaming
effect at radio band is dominated. The difference between $\rm
R_{*}$ and R, i.e. the correction of R, cover a wide range of
about three orders of magnitude. About half FSRQs have corrections
ranging from 10 to 100, and about one fifth of FSRQs ranging in
$100-1000$. In some extreme case, the correction is close to 1000.
However, $\rm R_{*}$ can be greater than R in a few FSRQs. This
implies that the optical luminosity in these sources are
significantly dominated by the extremely Doppler boosted
synchrotron emission resulting in the dominance of thermal
correction.
We note that for all FSRQs, the mean value of R is $\rm <R>=3.41$,
whereas the mean value of $\rm R_{*}$ is $\rm <R_{*}>=2.25$ after
correcting the effect of relativistic beaming in radio luminosity,
as well as taking the thermal corrected optical luminosity. By
adopting $\rm R_{*}$, some of radio loud quasars should in fact be
classified as radio intermediate and in some cases radio quiet.

\begin{figure}
\centerline{\epsfxsize=160mm\epsfbox{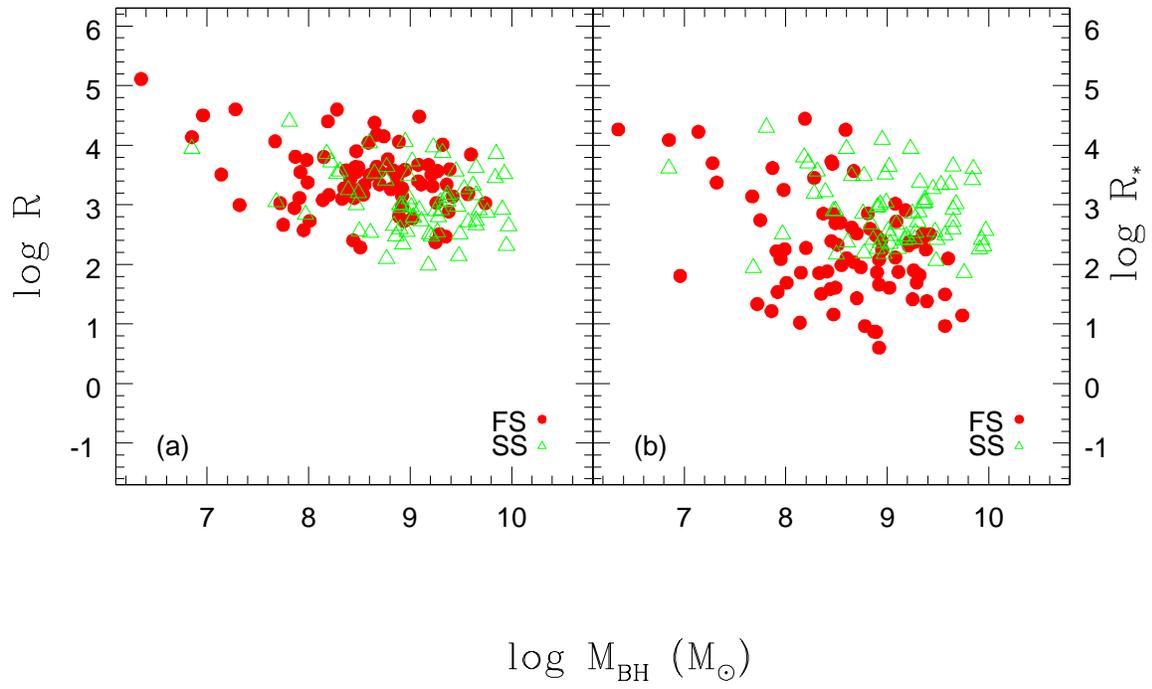}} \caption{(a) The
conventional radio loudness versus the black hole mass. (b) The
new-defined radio loudness versus the black hole mass. The symbols
are the same as in Fig. 5.}
\end{figure}

We explore the relationship between the black hole mass and radio
loudness R, and $\rm R_{*}$ in Fig. 8a, and Fig. 8b, respectively.
Apparently, a significant anti-correlation between black hole mass
and R presents in Fig. 8a. The Spearman rank correlation
coefficient is found to be $\rm r=-0.302$, with $\rm \ll0.01\%$
probability that no correlation is present. The detection of a
significant anti-correlation between black hole mass and radio
loudness is in good agreement with the Gu, Cao \& Jiang (2001)
study of a sample of radio loud quasars. However, Oshlack, Webster
\& Whiting (2002) attributed the apparent anti-correlation between
the radio loudness and black hole mass to the consequence of using
the optical flux in the measurement of black hole mass and also in
the calculation of radio loudness, for a sample of FSRQs.
Conversely, when the radio quiet AGNs are included, Lacy et al.
(2001) found a strong correlation on the combined FBQS+PG sample.
Moreover, Mclure \& Jarvis (2004) found similar results using a
sample of more than 6000 quasars from the SDSS, in which an upper
limit in radio loudness was calculated for those quasars
undetected by FIRST using the nominal FIRST object detection
threshold of 1 mJy. In contrast, Woo \& Urry (2002b) found no
indication of an R - $\rm M_{BH}$ correlation in their study of a
heterogeneous sample of 747 quasars in the redshift interval $\rm
1<z<2.5$ (although see McLure \& Jarvis (2004) for a different
interpretation).

However, we emphasize here that the conventional radio loudness R
is severely contaminated by the effects of relativistic beaming in
radio loud quasars as already shown in Fig. 7.
Strictly, it may not be right to simply use R to indicate the
radio property, with which further to investigate the relationship
with other parameters, e.g. black hole mass in present. Whenever
possible, the correction of beaming effect is required in
calculating the fundamental (or intrinsic) radio loudness for
FSRQs. Owing to the significance of corrections (as shown in Fig.
7), it's interesting to reinvestigate the correlation in Fig. 8a,
by replacing with the new-defined radio loudness, and to see how
the correction will influence the correlation.
Interestingly, the significant anti-correlation in Fig. 8a is not
present in Fig. 8b. We then come to the conclusion that the black
hole mass is not correlated with the new-defined radio loudness,
although it was with the conventional radio loudness, at least for
our sample of radio loud quasars. The significant anti-correlation
in Fig. 8a might be predominately caused by beaming effect in
radio luminosity and mistaking the beamed synchrotron emission
dominated optical luminosity, if not all. These results further
strengthen the importance of corrections in FSRQs, or radio loud
quasars in general.

To further tackle the importance and necessity of corrections in
radio loud quasars, we try to investigate the still debated issue
of radio loudness dichotomy, by combining our radio loud quasars
with a tentative sample of radio quiet quasars compiled from
literature. The black hole mass of radio quiet quasars are
estimated using the same methods as our radio loud quasars except
that their optical luminosity is directly used. Their conventional
radio loudness R are also calculated. The corrections in radio
loudness are not considered since it's not important in radio
quiet quasars. The distribution of conventional radio loudness R
and new-defined radio loudness $\rm R_{*}$ is shown in the upper
and lower panel of Fig. 9 for combined radio quiet and radio loud
quasars, respectively. It can be seen from the upper panel that
the gap around R=10, which is commonly used to distinguish radio
quiet and radio loud AGNs, is apparently present. Although our
combined sample is not complete, this is somehow consistent with
the so-called radio-loudness dichotomy found in the optically
selected quasars (Kellermann et al. 1989; Miller, Peacock \& Mead
1990; Ivezi\'{c} et al. 2002; also see Lacy et al. 2001 and
Cirasuolo et al. 2003 for an alternative viewpoint). In the lower
panel of Fig. 9, where the distribution of R of radio quiet
quasars and new-defined radio loudness $\rm R_{*}$ of radio loud
quasars are given, interestingly, we find that the radio loudness
became continuously distributed, and the gap around 10 in the
upper panel is not prominent. Expectedly, this result is simply
the consequence of corrections in radio loud quasars, which makes
enough radio loud quasars to fill in the gap.
This result strengthens the significance of corrections. Moreover,
it is likely suggestive that the uncorrections (both in radio and
optical luminosity) in radio loud quasars may be partly (if not
all) responsible for the observed radio loudness dichotomy in
optically selected quasars. Notably, the incompleteness of our
combined sample may preclude us to make solid conclusion.
Nevertheless, the fact that the apparent gap in radio loudness
distribution can be smoothed out after corrections, strongly
demonstrates the necessity of correction.

The apparent non-prominence of a gap in new-defined radio loudness
can also be clearly seen from the $\rm R_{*}-M_{BH}$ relation in
Fig. 10. Instead of no indication of correlation between the
new-defined radio loudness and black hole mass in radio loud
quasars alone (shown in Fig. 8b), we find a strong correlation in
the combined sample. The Spearman rank correlation coefficient is
found to be 0.451, at $\rm \gg$ 99.99 per cent confidence. This
result is in good agreement with that of McLure \& Jarvis (2004)
and Lacy et al. (2001). However, the scatter is large, and the
range in radio loudness at a given black hole mass is several
orders of magnitude. It's therefore clear that the influence of
other physical effect, such as the accretion rate, black hole spin
and source environment, must also be invoked to `unify' the radio
quiet and radio loud quasars.

\begin{figure}
\centerline{\epsfxsize=160mm\epsfbox{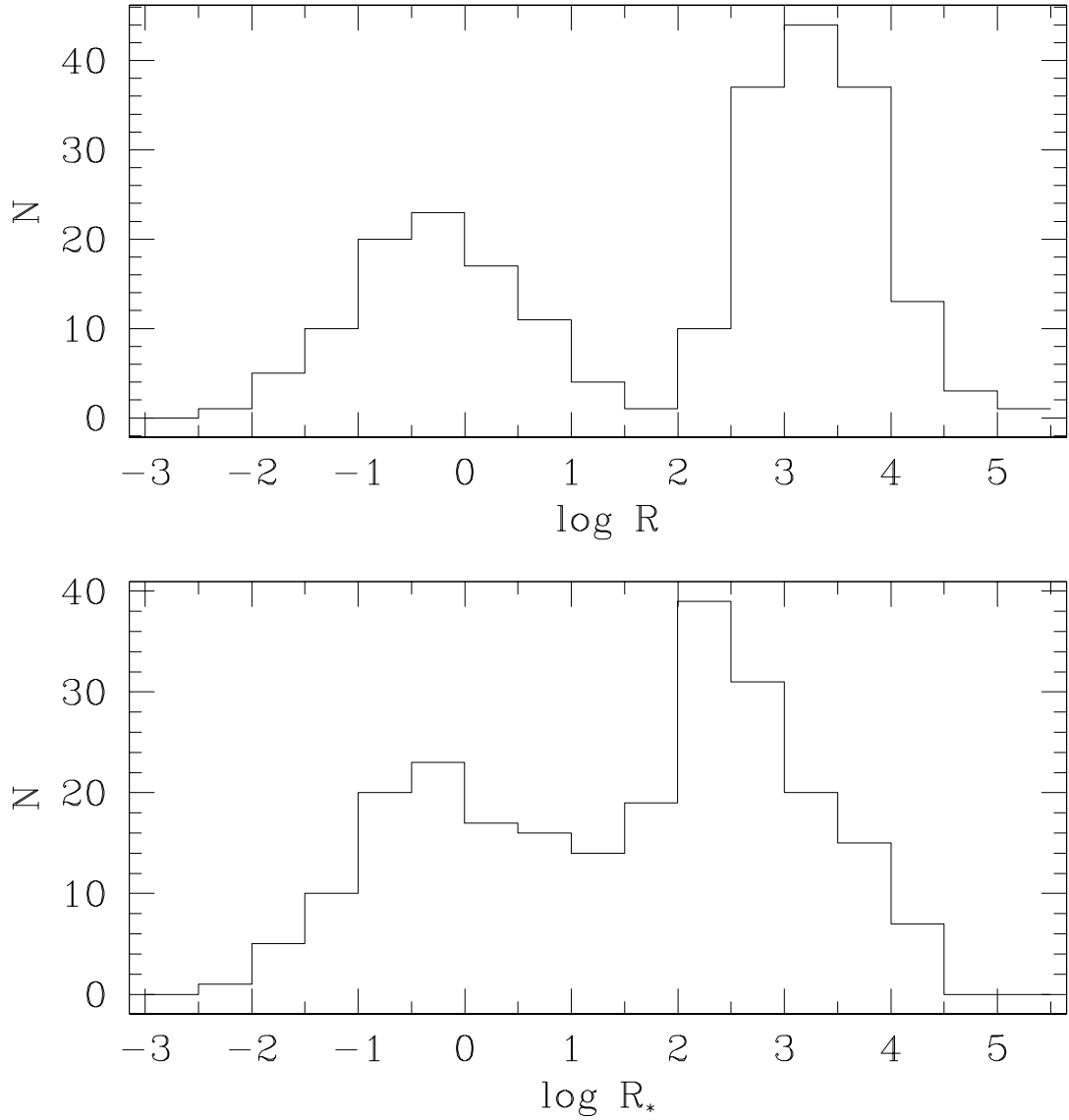}}
\caption{Distribution of the conventional radio loudness R (upper
panel) and the new-defined radio loudness $\rm R_{*}$ (lower
panel) for the combined sample of radio loud and radio quiet
quasars.}
\end{figure}

\begin{figure}
\centerline{\epsfxsize=160mm\epsfbox{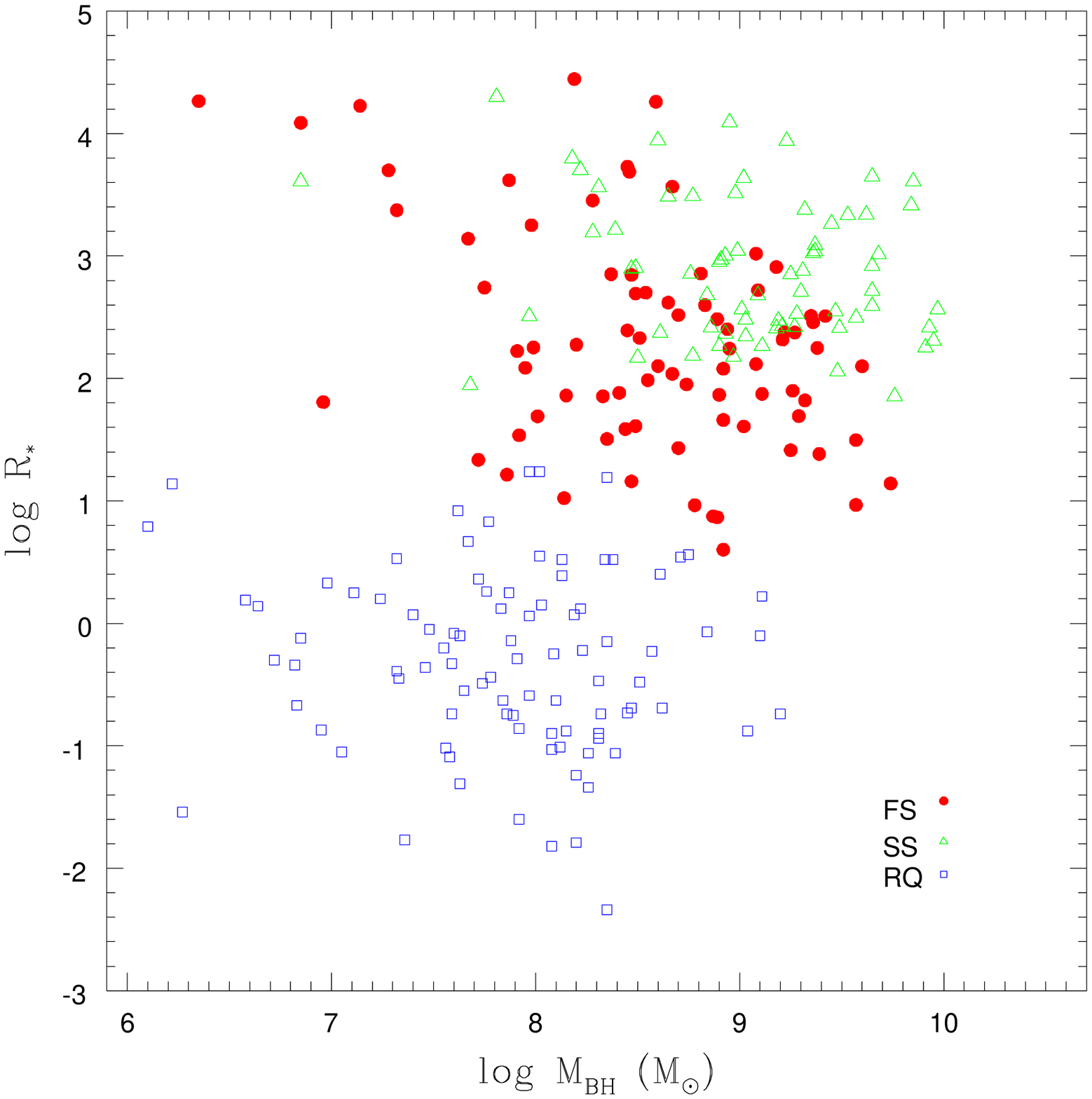}} \caption{$\rm
R_{*} - \rm M_{BH}$ relation for the combined sample of radio
quiet and radio loud quasars. Solid circles are for FSRQs. Open
triangles represent SSRQs, and open squares denote radio quiet
quasars.}
\end{figure}

\section{Conclusions}
By estimating the fundamental radio parameter, namely the jet
power using the extended radio emission, we have investigated the
relationship between the jet power and black hole mass
for a sample of 146 radio loud quasars compiled from literature.
Moreover, we define a new radio loudness as the ratio of radio
extended luminosity to the thermal optical luminosity estimated
from broad line luminosity. The relationship between the
new-defined radio loudness and black hole mass was investigated.
The main conclusions are summarized as follows:
\begin{itemize}
\item{After removing the effect of relativistic beaming in the
radio and optical emission, we find that the jet power is strongly
correlated with black hole mass for radio loud AGNs. This
correlation is proved to be intrinsic, and not an artefact of
common dependence of redshift.}

\item{When including the broad line region luminosity, we find
that the jet power is correlated with both black hole mass and
broad line region luminosity. The consistence with the Wang et al.
(2004) study indicates that this correlation is likely fundamental
in radio loud quasars. Moreover, we propose that the jet power
correlates with both black hole mass and Eddington ratio.}

\item{The correction in radio loudness is found to be significant
in FSRQs. The strong anti-correlation between the conventional
radio loudness and black hole mass is not present with new-defined
radio loudness.}

\item{Tentatively combining with a sample of radio quiet quasars,
we find that the apparent gap in the conventional radio loudness
is not prominent in the new-defined radio loudness. The necessity
of correction on radio loudness is emphasized. The black hole mass
is significantly correlated with the new-defined radio loudness in
the combined sample.}
\end{itemize}

\acknowledgements We are grateful to Xinwu Cao for helpful
discussion. We thank the anonymous referee for insightful comments
and constructive suggestions. This work is supported by NSFC under
grants 10373019, 10333020 and G1999075403. This research has made
use of the NASA/ IPAC Extragalactic Database (NED), which is
operated by the Jet Propulsion Laboratory, California Institute of
Technology, under contract with the National Aeronautics and Space
Administration.

{}

\end{document}